\documentclass[10pt,aps,prl,twocolumn,groupedaddress,showkeys]{revtex4-2}
\usepackage{amsmath}
\usepackage{graphicx}
\usepackage[colorlinks=true,linkcolor=blue,urlcolor=blue,citecolor=blue,anchorcolor=blue]{hyperref}
\DeclareGraphicsExtensions{.pdf,.eps,.png,.jpg,.mps}
\usepackage{epstopdf}
\usepackage{threeparttable}
\usepackage{color}
\usepackage[T2A]{fontenc}
\usepackage[utf8]{inputenc}
\usepackage{float}
\usepackage{tabularx}
\usepackage{array}
\usepackage[percent]{overpic}

\begin{document}

\title{Enhanced Terahertz Photoresponse via Acoustic Plasmon Cavity Resonances in Scalable Graphene}
\author{Domenico De Fazio$^{1,2,+}$}
\author{Sebastián Castilla$^{2,+}$}
\author{Karuppasamy P. Soundarapandian$^2$}
\author{Tetiana Slipchenko$^{3,4}$}
\author{Ioannis Vangelidis$^5$}
\author{Simone Marconi$^2$}
\author{Riccardo Bertini$^2$}
\author{Vlad Petrica$^6$}
\author{Yang Hao$^6$}
\author{Alessandro Principi$^7$}
\author{Elefterios Lidorikis$^5$}
\author{Roshan K. Kumar$^{2,8}$}
\author{Luis Martín-Moreno$^{3,4}$}
\author{Frank H. L. Koppens$^{2,9}$}
\email[]{domenico.defazio@unive.it}
\email[]{sebastian.castilla@icfo.eu}
\email[]{frank.koppens@icfo.eu}
\affiliation{$^1$ Department of Molecular Science and Nanosystems, Ca’ Foscari University of Venice, Via Torino 155, 30172 Venice, Italy}
\affiliation{$^2$ ICFO-Institut de Ciències Fotòniques, The Barcelona Institute of Science and Technology, Avinguda Carl Friedrich Gauss 3, 08860 Castelldefels (Barcelona), Spain}
\affiliation{$^3$ Instituto de Nanociencia y Materiales de Aragón (INMA), CSIC–Universidad de Zaragoza, 50009 Zaragoza, Spain}
\affiliation{$^4$ Departamento de F\'isica de la Materia Condensada, Facultad de Ciencias, Universidad de Zaragoza, 50009 Zaragoza, Spain}
\affiliation{$^5$ Department of Materials Science and Engineering, University of Ioannina, GR-451 10 Ioannina, Greece}
\affiliation{$^6$ School of Electronic Engineering and Computer Science, Queen Mary University of London, E1 4FZ, London, United Kingdom}
\affiliation{$^7$ Department of Physics and Astronomy, University of Manchester, Oxford Road, M13
9PL Manchester, United Kingdom}
\affiliation{$^8$ Catalan Institute of Nanoscience and Nanotechnology (ICN2), BIST and CSIC, Campus UAB, 08193 Bellaterra (Barcelona), Spain}
\affiliation{$^9$ ICREA - Instituci\'o Catalana de Recerca i Estudis Avan\c{c}ats, 08010 Barcelona, Spain}
\affiliation{$^+$ These authors contributed equally}


\begin{abstract}
Precise control and nanoscale confinement of terahertz (THz) fields are essential requirements for emerging applications in photonics, quantum technologies, wireless communications, and sensing. Here, we demonstrate a polaritonic cavity-enhanced THz photoresponse in an antenna-coupled device based on chemical-vapor-deposited (CVD) monolayer graphene. The dipole antenna lobes simultaneously serve as two gate electrodes, concentrate the impinging THz field, and efficiently launch acoustic graphene plasmons (AGPs), which drive a strong photo-thermoelectric (PTE) signal. Between 6 and 90$~$K, the photovoltage exhibits pronounced peaks, modulating the PTE response by up to $\sim 40\%$, that we attribute to AGPs forming a Fabry--P\'erot THz cavity in the full or half graphene channel. Combined full-wave and transport–thermal simulations accurately reproduce the gate-controlled plasmon wavelength, spatial absorption profile, and the resulting nonuniform electron heating responsible for the PTE response. The lateral and vertical maximum confinement factors of the AGP wavelength relative to the incident wavelength are 165 and 4000, respectively, for frequencies from 1.83 to 2.52 THz. These results demonstrate that wafer-scalable CVD graphene, without hBN encapsulation, can host coherent AGP resonances and exhibit an efficient polaritonic-enhanced photoresponse under appropriate gating, antenna coupling, and AGP cavity design, opening a route to scalable, polarization and frequency-selective, liquid-nitrogen cooled, and low-power consumption THz detection platforms based on plasmon–thermoelectric transduction.
\end{abstract}

\maketitle

\section{Introduction} \label{Intro}
Far-infrared or terahertz (THz) radiation, spanning wavelengths between 15–1000 \textmu m, is becoming increasingly important across disciplines due to its unique ability to probe matter non-invasively and with high spectral specificity~\cite{Dhillon2017}. It has found applications in biomedical imaging~\cite{Yang2016}, security screening~\cite{Li2023}, high-speed wireless communications~\cite{Rappaport2019}, and chemical sensing~\cite {Walther2010}, leveraging its non-ionizing nature and ability to interact with low-energy excitations in materials. The growing demand for THz technology has spurred significant advancements in both photonic~\cite{Mittleman2018, Nagatsuma2016} and electronic~\cite{Rieh2020, Sengupta2018} device platforms.

A variety of technologies have been developed to detect THz radiation~\cite{Lewis2019, Freeman2025}, each with strengths and limitations in terms of sensitivity, speed, and complexity. Among the most common ones, Schottky diode detectors offer room-temperature operation and fast ($\sim$ns) response~\cite{Song2012, Koenig2013, Rettich2015, Mehdi2017}, but their sensitivity degrades significantly above 1--2~THz~\cite{Freeman2025}. Quantum well infrared photodetectors (QWIPs) and quantum cascade detectors can provide fast, frequency-selective detection~\cite{Liu2005, Li2017, Yachmenev2022} but are constrained by narrow bandwidth and the need for cryogenic operation~\cite{Lewis2019}. Thermal detectors, including cryogenically cooled superconducting bolometers, uncooled microbolometers, Golay cells, thermopiles, and more, are among the most sensitive and broadband THz detectors~\cite{Lewis2019, Bielecki2024} but are typically slow ($\sim$ms) and often require thermal isolation or active cooling~\cite{Zhang2022, Vicarelli2022}.

Graphene offers a compelling platform for THz detection due to the combination of broadband absorption, high carrier mobility, electrostatic tunability, low specific heat, and weak electron-phonon coupling~\cite{CastroNeto2009, Falkovsky2008, Zihlmann2019}. As a gapless two-dimensional semimetal, graphene supports interaction with THz radiation across a wide frequency range, while encapsulation with the atomically flat hBN preserves mobility and screens substrate scattering~\cite {Cai2014, Bandurin2018}. Detection schemes such as the photo-thermoelectric effect (PTE)~\cite{Cai2014, Castilla2019, Ludwig2024}, Dyakonov-Shur (DS) rectification~\cite{Vicarelli2012, Soltani2020}, ballistic transport~\cite{Auton2017}, and tunneling~\cite{Gayduchenko2021, Mylnikov2023} have been experimentally demonstrated to exhibit broadband and fast response at room and cryogenic temperatures~\cite{Viti2021, Agarwal2023}. However, graphene's atomically thin nature limits free-space absorption, requiring engineered strategies to enhance light-matter coupling~\cite{Liu2018}.

Polaritons offer a powerful strategy to compress electromagnetic waves far below the free-space wavelength and enhance light-matter interactions. These hybrid light-matter quasiparticles, including plasmon polaritons in graphene~\cite{Grigorenko2012, Castilla2024}, phonon polaritons in hexagonal boron nitride (hBN)~\cite{Dai2014, Castilla2020, Castilla2024}, and more~\cite {Xie2025, Abajo2025, Lazzari2020}, can confine incident radiation to deeply sub-diffractional volumes, greatly increasing local field intensity and absorption~\cite{Basov2016, Low2017}. Most recently, \textit{acoustic} graphene plasmons (AGPs) have emerged as a promising alternative to \textit{conventional} (optical) plasmons: by coupling a graphene layer to a nearby metal or another graphene sheet through a dielectric spacer, the system can host anti-symmetric charge oscillations that result in even tighter (THz) field confinement, and enhanced absorption~\cite{Principi2011, Alcaraz2018, Epstein2020, Barcons2023, Castilla2024}. These acoustic modes offer new opportunities for efficient THz detection by pushing interaction volumes to the extreme nanoscale~\cite{Epstein2020}.

Several experimental studies have leveraged conventional and acoustic plasmons in graphene THz photodetectors and photocurrent spectroscopy architectures~\cite{Cai2015, Alonso2017, Bandurin2018, Caridad2024}. For example, Cai \textit{et al.} demonstrated that patterning graphene into microribbons can introduce a tunable plasmonic resonance at THz frequencies, thereby increasing absorption and enabling PTE THz detection at room temperature~\cite{Cai2015}. Alonso-González \textit{et al.} directly visualized acoustic plasmons in a split-gated graphene photodetector by near-field photocurrent mapping, confirming the expected $\lambda_\mathrm{p} \ll \lambda_0$ (approximately $\lambda_0/66$) and the linear dispersion of AGPs due to coupling with the metal gate~\cite{Alonso2017}. Bandurin \textit{et al.} achieved resonant far-field THz detection by using a high-mobility hBN-encapsulated bilayer graphene channel as a plasmonic Fabry--P\'erot cavity, observing multiple gate-tunable photocurrent peaks when the channel length accommodated an integer number of plasmon half-waves~\cite{Bandurin2018}. Notably, the resonances were visible only under cryogenic conditions ($\sim$10 K), whereas at room temperature or lower THz frequencies, the plasmonic resonances were overdamped and faded in the noise. In a recent study, a graphene-based THz detector demonstrated plasmonic resonances up to room temperature\cite{Caridad2024}. However, these resonances appeared only as weak features in the photoresponse\cite{Caridad2024} and therefore did not demonstrate their full potential for the aforementioned applications.

The above-mentioned works underscore that achieving a far-field resonant AGP THz response in scalable CVD single-layer graphene, without hBN encapsulation, has not yet been realized. In addition, an efficient method to exploit graphene plasmons to enhance the far-field photoresponse in the THz range remains elusive. These challenges motivate the exploration of such a platform for AGP-enhanced THz detectors.

In this work, we demonstrate resonant polaritonic cavity-enhanced THz photoresponse in a split-gated architecture based on chemical vapor deposited (CVD) single-layer graphene. The gate electrodes are shaped to function as a dipole antenna, designed to efficiently couple 1.83 to 2.52~THz radiation into the graphene channel, act as an AGP launcher, and asymmetrically dope graphene to harness the PTE effect. When cooled to cryogenic temperatures (down to 5~K), we observe pronounced resonances that strongly modulate the photovoltage response as a function of the gate voltage, consistent with AGPs forming standing wave modes in the device. The observed features indicate Fabry--P\'erot-type resonances both across the full graphene channel and localized beneath one of the two gates (half channel). These resonances vanish at 130~K while warming the sample up. 

So far, graphene plasmonic resonances have typically appeared only as small perturbations in the photovoltage, at least one to two orders of magnitude smaller than the conventional photoresponse\cite{Caridad2024, Bandurin2018}. Here, we overcome this issue and demonstrate a plasmonically enhanced photoresponse that is $30\%$ higher than the maximum \textit{conventional} PTE response. We have designed the AGP resonances to align closely in Fermi energy with the peak of the Seebeck coefficient to maximize the polaritonic cavity-enhanced photoresponse. This shows that our THz AGP cavity geometry is efficiently exploited to enhance the photoresponse and paves the way for further improvements. In contrast to the DS mechanism, in which resonances occur along the entire channel\cite{Bandurin2018, Caridad2024}, our cavity geometry defines the AGP resonance within the region where the antenna lobe overlaps with the graphene channel. 

The lateral and vertical maximum confinement factors\cite{Castilla2024, Alcaraz2018, Lee2019} of the AGP wavelength relative to the incident wavelength are 165 and 4000, respectively, for frequencies ($f$) from 1.83 to 2.52~THz. Our findings establish that CVD-grown graphene, without the need for hBN encapsulation, can support coherent AGP resonances under suitable architectures, gating conditions, and liquid-nitrogen cooling. By further improving the mobility of CVD graphene through novel growth approaches\cite{Gebeyehu2024, Li2022}, one can expect the AGP resonances to be maintained up to room temperature. This opens a pathway toward scalable, resonantly enhanced THz photodetectors based on AGPs, with potential applications in polarization- and frequency-selective imaging, spectroscopy, communications, and low-power on-chip THz sensing.

\section{Results and Discussion} \label{Results}

\subsection{Device Concept and Simulations} \label{Device}

\begin{figure*}[htbp!]
\centerline{\includegraphics[width=180mm]{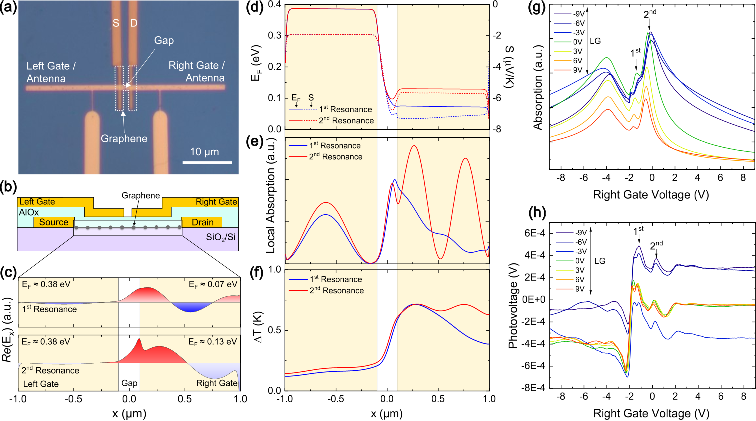}}
\caption{
(a) Optical microscope top-view image of the split-gated CVD graphene device. The left and right gates are shaped as a dipole antenna with a $200$~nm gap, while the graphene channel (H-shaped, dashed region) connects the source (S) and drain (D) electrodes.  
(b) Schematic cross-sectional view of the device, showing the AlO$_x$ gate dielectric, graphene channel, SiO$_2$/Si substrate, and the split-gate configuration.  
(c) Simulated spatial distribution of the real part of the in-plane electric field, $\mathrm{Re}(E_x)$, along the graphene channel. The results are shown for the representative left-gate voltage value of $9~\text{V}$ and two right-gate voltages, $-1.6~\text{V}$ and $-0.6~\text{V}$, labeled as the 1$^{\mathrm{st}}$ and 2$^{\mathrm{nd}}$ resonance, respectively. $T = 6$ K, as in (d-g).
(d) Simulated Fermi energy distribution $E_F(x)$ (solid lines) and Seebeck coefficient $S(x)$ (dotted lines) across the channel for the first (blue) and second (red) gating voltage, \textit{i.e.}, $-1.6~\text{V}$ and $-0.6~\text{V}$, highlighting the carrier density modulation induced by asymmetric gating.
(e) Calculated local optical absorption profile for the same gate voltages, showing enhanced absorption in the gap region and under the right gate.  
(f) Simulated electronic temperature increase $\Delta T$ along the channel for the two resonances, illustrating localized hot-electron formation in regions of high field confinement.  
(g) Simulated total optical absorption as a function of right-gate voltage for several fixed left-gate (LG) voltages, revealing gate-tunable AGP resonances (the bidirectional arrow highlights the used LGs). Illumination frequency is 2.5~THz. The electron and hole mobilities under the left and right antenna gates were chosen in agreement with values extrapolated from fits of experimental data in Supplementary Fig.~S1; under the antenna gap, the mobility was assumed to vary linearly between these values.
(h) Experimentally measured photovoltage at 2.5~THz and 18~mW incident power as a function of right-gate voltage at several fixed left-gate (LG) voltages, showing resonant features consistent with AGP Fabry--P\'erot standing waves in the graphene channel.
}
\label{fig:Figure1}
\end{figure*}

Fig.~\ref{fig:Figure1}(a) shows a microscope top-view image of our device, while Fig.~\ref{fig:Figure1}(b) is a sketched side view. Details of the device fabrication steps are discussed in the \hyperref[Sec:Exp]{Experimental Section}. CVD monolayer graphene was grown on copper foil and subsequently wet-transferred onto a highly resistive silicon (Si) substrate ($\rho > 10,000~\Omega\cdot$cm) with a $\sim$300 nm thermally grown silicon dioxide (SiO$_2$) layer. The source (S) to drain (D) electrode separation, \textit{i.e.}, the graphene channel length, is $L\sim2~\mu$m, matching the channel width of $W\sim2~\mu$m. A $\sim$40 nm-thick aluminum oxide (AlO$_x$) layer, deposited by atomic layer deposition (ALD), serves as the gate dielectric. The graphene was patterned into an H-shaped channel using reactive ion etching, a geometry chosen to optimize the contact resistance between graphene and the metallic electrodes~\cite{Viti2021}. The split-gate electrodes, designed as the two symmetric half-lobes of a dipole antenna, serve a dual role: they act as metallic dipole arms resonant near 2.5~THz and simultaneously define the graphene channel cavity where acoustic graphene plasmons (AGPs) form. The dipole arms are designed to approximate a half-wavelength resonance at 2.5~THz, yielding a capacitive near-field impedance that couples efficiently to the low-impedance plasmonic mode in graphene. In this configuration, the antenna converts incident far-field THz radiation into a strongly localized electric field concentrated across the dipole antenna gap, providing efficient near-field excitation of the AGP modes. As described in the following sections, the spatial overlap between the antenna lobes and the graphene channel defines the AGP THz cavity. The gap between the two gates/antenna lobes is 200~nm, chosen based on the resolution limits of our electron-beam lithography system and to avoid leakage between the antenna branches. The total antenna length is 40~$\mu$m, optimized through full-wave electromagnetic simulations (see Supplementary Material, Fig.~S2). This design enables simultaneous optical coupling and electrostatic control of the graphene channel, facilitating the exploitation of the PTE effect~\cite{Castilla2019}. 

The simulated spatial distribution of the real part of the in-plane electric field, $\mathrm{Re}(E_x)$, shown in Fig.~\ref{fig:Figure1}(c), corresponds to a gating configuration used in the experiment: the left gate (LG) is kept at $V_{\mathrm{LG}} = 9~\text{V}$, while the right-gate (RG) voltage $V_{\mathrm{RG}}$ is swept. We show results for two values, $V_{\mathrm{RG}} = -1.6~\text{V}$ and $V_{\mathrm{RG}} = -0.6~\text{V}$, which coincide with the two experimentally-observed photovoltage peaks hereafter labeled as the 1$^{\mathrm{st}}$ and 2$^{\mathrm{nd}}$ resonances, respectively. In both panels, the yellow shading marks the regions beneath the metallic gates, while the central white area denotes the $200~\text{nm}$ antenna gap in the $L = 2~\mu\text{m}$-long graphene channel.
At the 1$^{\mathrm{st}}$ resonance (top panel), the field oscillations extend across the entire graphene channel. The distribution exhibits a sequence of nodes and antinodes forming a clear standing-wave pattern. The resonance profile is characterized by two negative lobes (blue, $\mathrm{Re}(E_x) < 0$), one strong positive lobe (red, $\mathrm{Re}(E_x) > 0$), and a truncated positive half-lobe near the right boundary, indicating a slight deviation from an ideal symmetric standing wave. The smooth evolution of the field from the left gated region, through the gap, and into the right gated region indicates that the full channel length acts as the cavity, \textit{i.e.}, the cavity length $L_{\mathrm{c}} = L$. Using the Fabry--P\'erot resonance condition for nearly ideal mirrors,

\begin{equation}
    \lambda_{\mathrm{p}}^{\mathrm{eff}} = \frac{2L_{\mathrm{c}}}{m},
    \label{eq:FP}
\end{equation}

with $L_{\mathrm{c}} = L$ and mode index $m = 4$, we obtain an \textit{effective} plasmon wavelength $\lambda_{\mathrm{p}}^{\mathrm{eff}} \approx L/2$. The strong AGP reflection at the gold contacts justifies the assumption of nearly ideal mirrors.
In contrast, at the 2$^{\mathrm{nd}}$ resonance (bottom panel), the field distribution exhibits a qualitatively different behavior. The oscillation pattern beneath the left gate closely resembles that of the 1$^{\mathrm{st}}$ resonance, as expected, since in both cases the left region is tuned to the same Fermi energy $E_{\mathrm{F}}$ by $V_{\mathrm{LG}} = 9~\text{V}$. However, the overall field profile is markedly different: we observe an abrupt transition in the gap region and truncated field amplitudes near both contacts. This abrupt change originates from a sharp variation in carrier density---and thus surface conductivity---within the gap region (see the $E_{\mathrm{F}}$ distributions in Fig.~\ref{fig:Figure1}(d)), which leads to strong plasmon backscattering, in agreement with previous work~\cite{Garcia-Pomar2013}. By contrast, for the 1$^{\mathrm{st}}$ resonance, the conductivity profile varies more smoothly across the gap, so that AGP reflection at this interface is strongly reduced and plasmon backscattering is suppressed.
In the 2$^{\mathrm{nd}}$-resonance configuration, the abrupt conductivity change at the gap edge, together with the metal contacts, effectively turns the graphene channel into two sub-cavities of length $L_{\mathrm{c}} = L/2$, each bounded by a metal contact at one end and by the gap region at the other. In principle, each sub-cavity can support its own plasmonic modes. In the right sub-cavity, the field develops an approximate node--antinode--node--antinode sequence, which can again be described by the Fabry--P\'erot condition.

\begin{equation}
    \lambda_{\mathrm{p},R} = \frac{2L_{\mathrm{c}}}{m},
    \label{eq:FP-half}
\end{equation}

with $L_{\mathrm{c}} = L/2$ and $m = 2$, yielding $\lambda_{\mathrm{p},R} \approx L/2$. This appears as alternating red and blue lobes emerging from the gap edge. The similarity of the oscillatory pattern under the left gate at both resonances, together with the clear formation of two sub-cavities, indicates that the experimentally observed 2$^{\mathrm{nd}}$ photovoltage peak is primarily associated with a Fabry--P\'erot AGP mode confined in the right sub-cavity.
Altogether, these oscillatory field profiles provide direct evidence of Fabry--P\'erot--type AGP modes, with the 1$^{\mathrm{st}}$ resonance corresponding to a mode extending across the full channel and the 2$^{\mathrm{nd}}$ resonance corresponding to a mode localized mainly beneath the right gate. The distinct behavior observed at the two right-gate voltages will be analyzed in more detail in the following sections.

The oscillatory field distributions in Fig.~\ref{fig:Figure1}(c) originate from the asymmetric carrier-density profile along the channel. By applying different voltages to the left and right gates, a non-uniform carrier concentration profile $n(x)$ is induced in the graphene channel. Since the Fermi energy is directly related to the local carrier density through $E_{\mathrm{F}}(x) = \text{sgn}\!\big(n(x)\big)\hbar v_{\mathrm{F}} \sqrt{\pi |n(x)|}$, the gating asymmetry results in significantly different $E_{\mathrm{F}}(x)$ values beneath the two gates. The corresponding Fermi-energy distributions are shown in Fig.~\ref{fig:Figure1}(d) for the same gate-voltage configurations as in Fig.~\ref{fig:Figure1}(c).  The spatial modulation of $E_{\mathrm{F}}(x)$ directly determines the local optical absorption of the graphene channel, which can be expressed as $A(x) \propto \mathrm{Re}[\sigma(E_{\mathrm{F}}(x))]\;|E_x(x)|^2$, where $\sigma(E_{\mathrm{F}}(x))$ is the graphene optical conductivity, which is spatially dependent on the Fermi energy. The local absorption is illustrated in Fig.~\ref{fig:Figure1}(e) for the two gate-voltage configurations discussed above. For both resonances, absorption is strongly enhanced inside the antenna gap. For the 1$^{\mathrm{st}}$ resonance (blue curve), the gap peak is followed by a weaker extension beneath the right gate and a smaller, smoother peak beneath the left antenna. In contrast, for the 2$^{\mathrm{nd}}$ resonance (red curve), the maximum absorption peak is larger, with the gap peak followed by multiple pronounced maxima under the right gate and a weaker one beneath the left gate. These spatially localized absorption hotspots provide the direct source of electronic heating: the absorbed optical power is dissipated into the electronic system, producing the nonuniform electron-temperature profiles $T_{\mathrm{e}}(x)$ shown in Fig.~\ref{fig:Figure1}(f)~\cite{Castilla2020}. The local temperature rise relative to equilibrium, $\Delta T(x) = T_{\mathrm{e}}(x) - T$, closely follows the absorption pattern. For the 1$^{\mathrm{st}}$ resonance (blue), the temperature rise is sharply peaked just to the right of the gap. For the 2$^{\mathrm{nd}}$ resonance (red), the heating profile is broader and more extended, with maxima shifted farther into the gated region. This spatially varying $T_{\mathrm{e}}(x)$ generates a thermoelectric voltage according to $V_{\mathrm{PTE}} \propto S(x)\,\nabla T_{\mathrm{e}}(x)$, where $S(x)$ denotes the Seebeck coefficient determined by the local Fermi energy through the Mott relation~\cite{Zuev2009}:
\begin{equation}
    S(x) = - \frac{\pi^2 k_{\mathrm{B}}^2 T_{\mathrm{e}}(x)}{3e} \, \frac{1}{\sigma(E_{\mathrm{F}}(x))} \, \frac{\partial \sigma(E_{\mathrm{F}}(x))}{\partial E_{\mathrm{F}}(x)} ,
    \label{eq:Mott}
\end{equation}
where $k_{\mathrm{B}}$ is the Boltzmann constant, $e$ the elementary charge, $\sigma(x)$ the local electrical conductivity, and $E_{\mathrm{F}}(x)$ the local Fermi energy. For the quantitative simulations of the $T_{\mathrm{e}}(x)$ profiles, we used our full thermoelectric framework, in which $S(x)$ is evaluated in its energy-resolved form via Boltzmann/Onsager transport (see Methods of Ref.~\citenum{Vangelidis2022}). The spatial variation of $S(x)$ is shown by the dotted curves in Fig.~\ref{fig:Figure1}(d). The values of $S(x)$ are smaller at cryogenic temperatures compared to those at room temperature, as expected from the Mott relation~\cite{Duan2016}.

It is well established that the response of the photodetector can be captured by the simulated total absorption of graphene~\cite{Bylinkin2024, Castilla2020}. In Fig.~\ref{fig:Figure1}(g), we present the calculated total absorption (integral of local absorption across the entire channel) of graphene for several fixed left-gate voltages while tuning the right-gate voltage. The spectra exhibit three prominent absorption peaks: two on the electron-doped side of the right gate charge neutrality point (CNP, right side), whose positions remain nearly invariant with $V_{\mathrm{LG}}$, and one on the hole-doped side (left side). The CNP is assumed to be at $V_{\mathrm{RG}}\sim$ -2 V for the right gate and at $V_{\mathrm{LG}}\sim$ -4 V for the left gate, as per the experimental measurements. The difference in the CNP found in the experiments when sweeping each of the gates is likely due to fabrication process residues and local inhomogeneities. Furthermore, previous work has shown that ALD AlO$_x$ on graphene can drive the material into $n$-doping, by virtue of charge transfer at the graphene/oxide interface~\cite{Riazimehr2025}. The peak amplitudes depend strongly on $V_{\mathrm{LG}}$: more negative left-gate voltages (blue curves) result in stronger absorption, whereas more positive left-gate voltages (yellow curves) suppress the response. All three peaks originate from AGP mode excitations. For $V_{\mathrm{LG}}=9$~V, the first and second resonances on the electron-doped side are located exactly at $V_{\mathrm{RG}}=-1.6$~V and $V_{\mathrm{RG}}=-0.6$~V, respectively. For other left-gate voltages, the resonance positions remain very close to these values. This explains the choice of right-gate voltages used for the profiles shown in Fig.~\ref{fig:Figure1}(c–f).

\subsection{Optoelectronic Measurements} \label{Opto}

Figure~\ref{fig:Figure1}(h) shows the measured photovoltage $V_{\mathrm{ph}}$ as a function of the right-gate voltage $V_{\mathrm{RG}}$ for several fixed values of the left-gate voltage $V_{\mathrm{LG}}$ at a temperature of $T=6$~K. The data reveal a rich structure dominated by a prominent polarity inversion near the graphene CNP and multiple resonant features that we attribute to the excitation of acoustic AGPs. A sharp sign reversal of the photovoltage occurs around $V_{\mathrm{RG}} \approx -2$~V, which corresponds to the CNP of the right-gated graphene region. This feature, together with the strong peak at $V_{\mathrm{RG}}$ just above $V_{\mathrm{CNP}}$ (\textit{i.e.}, at $\approx -1.6$~V), followed by a slowly decaying tail, is a hallmark of the PTE effect in graphene-based devices~\cite{Castilla2019, Gabor2011}. This peak in photovoltage, $V_{\mathrm{ph}}$, arises from the steep variation of $S$ with carrier density between the two asymmetrically gated graphene regions under the two branches of the dipole antenna, combined with the large photo-induced $\Delta T$ localized in the antenna gap.

\begin{figure*}[htbp!]
\centerline{\includegraphics[width=180mm]{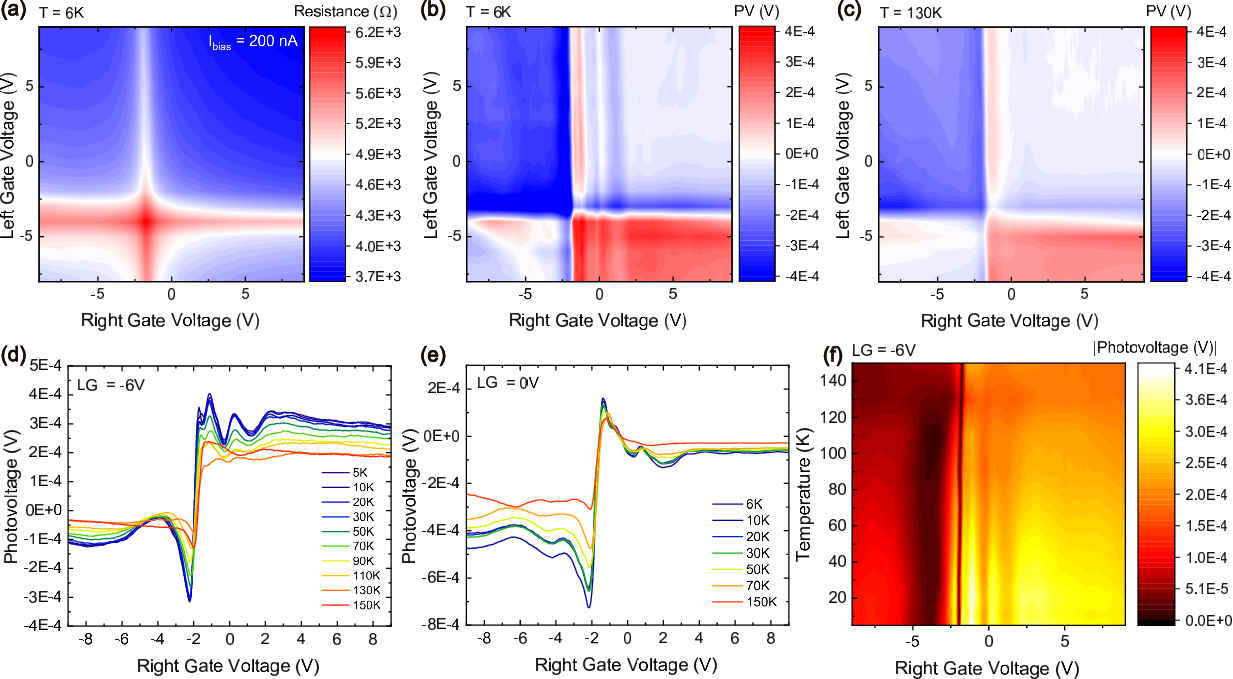}}
\caption{
Gate- and temperature-dependent resistance, photovoltage, and AGP resonances.
(a) Two-dimensional resistance map as a function of left- and right-gate voltages at $T=6$~K, measured with $I_{\mathrm{bias}}=200$~nA. 
(b) Corresponding photovoltage $V_{\mathrm{ph}}$ map at $T=6$~K under $f=2.5$~THz illumination, showing the characteristic sixfold polarity pattern of the PTE effect and sharp resonant features associated with AGPs.  
(c) Same measurement at $T=130$~K, where AGP resonances are strongly suppressed by increased plasmon damping.  
(d,e) Line cuts of $V_{\mathrm{ph}}$ vs $V_{\mathrm{RG}}$ for several temperatures at fixed $V_{\mathrm{LG}}=-6$~V (d) and $V_{\mathrm{LG}}=0$~V (e), revealing two prominent AGP resonances around $V_{\mathrm{RG}}\approx-1.2$~V and $0.3$~V that gradually weaken with increasing $T$ and are enhanced when Seebeck asymmetry is maximized.  
(f) False-color map of $|V_{\mathrm{ph}}|$ vs $V_{\mathrm{RG}}$ and temperature at $V_{\mathrm{LG}}=-6$~V, highlighting the progressive damping and slight redshift of the AGP resonances with increasing $T$.}
\label{fig:Figure2}
\end{figure*}

At gate voltages above the CNP, two distinct resonant peaks are visible in Fig.~\ref{fig:Figure1}(h) at $\approx-1.2$~V and $\approx0.3$~V, labeled as the 1$^{\mathrm{st}}$ and 2$^{\mathrm{nd}}$ resonance, respectively. These are very close to the theoretically-predicted values of $\approx-1.6$~V and $\approx-0.6$~V, and correspond to the standing-wave modes of AGPs predicted by simulations. Their positions evolve moderately with $V_{\mathrm{LG}}$, indicating that the resonances are primarily controlled by the carrier density under the right gate, while the left gate mostly tunes the global carrier asymmetry required for PTE detection. These peaks coincide with enhanced local absorption in the graphene channel, as confirmed by the simulated absorption profile in Fig.~\ref{fig:Figure1}(e), demonstrating a direct link between plasmon-enhanced field confinement and the observed photovoltage response. In addition to these two well-pronounced features, a third resonance appears at V$_{\mathrm{RG}} <$ CNP. This resonance is discernible in the experimental spectra but is significantly weaker than predicted by the simulations. The origin of this reduced visibility is not fully clear, but it may be related to increased disorder or other asymmetries under negative bias that are not captured by the model. Furthermore, it is important to distinguish the \textit{antenna} resonance from the \textit{plasmonic cavity} resonances: while the metallic dipole antenna provides broadband coupling around its fundamental resonance ($\sim$2.5~THz), the pronounced photovoltage peaks in Fig.~\ref{fig:Figure1}(h) arise from AGP Fabry–-Pérot resonances confined within the channel. In this picture, the antenna primarily serves as an efficient launcher and receiver of these localized plasmons, rather than setting the overall device resonance. In the following, we therefore focus our analysis on the two resonances above the CNP, which provide the clearest and most reproducible signatures of AGPs.

The dependence on $V_{\mathrm{LG}}$ also reveals subtle features: at increasingly negative $V_{\mathrm{LG}}$ (from -3 to -9 V, blue traces), the overall magnitude of $V_{\mathrm{ph}}$ increases, consistent with stronger PTE contributions arising from larger Seebeck asymmetry between the two gated regions. At the same time, several smaller satellite peaks appear at higher carrier densities. These may originate from higher-order plasmonic overtones or weak cavity-assisted effects; however, their amplitudes are much lower than the primary 1$^{\mathrm{st}}$ and 2$^{\mathrm{nd}}$ AGP resonances, and thus they are not analyzed in this study.

\begin{figure*}[htbp!]
\centerline{\includegraphics[width=180mm]{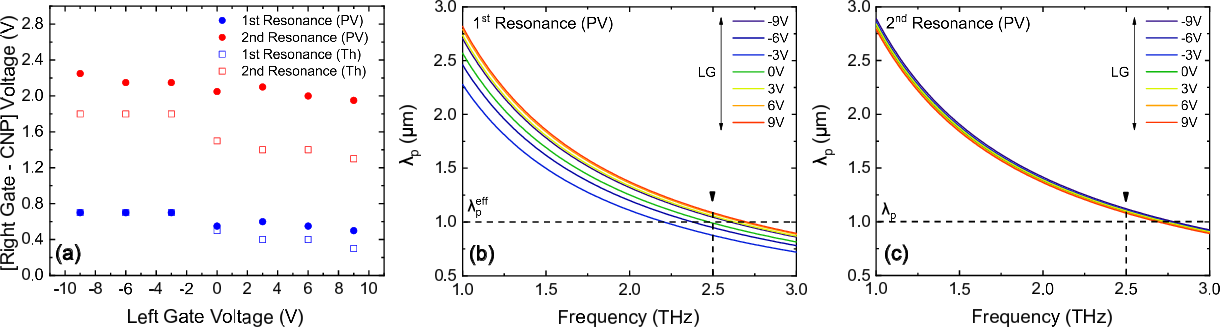}}
\caption{
(a) Comparison between measured photovoltage (PV/filled symbols) and simulated absorption (Th/open symbols) peak positions for the first and second AGP resonances in the split-gated graphene device. Blue and red symbols correspond to the first and second resonances, respectively.  
(b,c) Colored curves correspond to the AGP dispersions calculated at the right-gate voltages corresponding to the first (b) and second (c) resonances extracted from the photovoltage spectra. Horizontal dashed lines mark the Fabry--P\'erot conditions: $\lambda_{\mathrm{p}}^{\mathrm{eff}} = L/2$ for a mode spanning the full channel (b), and $\lambda_{\mathrm{p}} = L/2$ for a mode confined mainly to the right half of the channel, where $\lambda_{\mathrm{p}}$ denotes the local plasmon wavelength under the right gate (c). Vertical dashed lines with arrows indicate the experimental resonance frequency at $\sim 2.5~\text{THz}$.
}
\label{fig:Figure3}
\end{figure*}

Fig.~\ref{fig:Figure2} provides a comprehensive view of the gate-dependent resistance and photovoltage response, highlighting the interplay between PTE effects and resonant excitation of AGPs. Fig.~\ref{fig:Figure2}(a) shows the two-dimensional resistance map $R(V_{\mathrm{LG}}, V_{\mathrm{RG}})$ measured at $T=6$~K using a constant source-drain bias current of $I_{\mathrm{bias}} = 200$~nA (see \hyperref[Sec:Exp]{Experimental Section} for details on the measurement setup). The map exhibits the characteristic four-quadrant symmetry typical of dual-gated graphene devices\cite{Castilla2019}. The central bright region of enhanced resistance corresponds to the CNP region, where the carrier density beneath the two gates approaches zero. The CNP is at $V_{\mathrm{RG}}\sim$ -2 V and $V_{\mathrm{LG}}\sim$ -4 V. Moving away from the CNP along either gate axis, $R$ decreases due to electrostatic doping of the graphene. Next, we illuminate the device with a continuous-wave 2.5~THz laser at $T=6$~K, and display the corresponding photovoltage map $V_{\mathrm{ph}}(V_{\mathrm{LG}}, V_{\mathrm{RG}})$ in Fig.~\ref{fig:Figure2}(b). Unlike the \textit{fourfold} symmetry of $R$, the photovoltage shows a \textit{sixfold} pattern, with six alternating regions of positive (red) and negative (blue) photovoltage values. This six-quadrant symmetry is another hallmark of the PTE effect in graphene-based THz photodetectors~\cite{Lemme2011, Gabor2011, Castilla2019}, arising from the opposite signs of the Seebeck coefficient $S$ across different carrier-type configurations. Superimposed on this sixfold structure are additional finer modulations of $V_{\mathrm{ph}}$, seen as oscillatory ripples running parallel to the $V_{\mathrm{RG}}$ axis in Fig.~\ref{fig:Figure2}(b). These correspond to the resonant excitation of AGPs, in agreement with simulations in Fig.~\ref{fig:Figure1}(c-g), and consistent with the sharp peaks observed in Fig.~\ref{fig:Figure1}(h) at $V_{\mathrm{RG}} \approx -1$~V (1$^\mathrm{st}$ resonance) and $V_{\mathrm{RG}} \approx 0.3$~V (2$^\mathrm{nd}$ resonance). The comparison between Figs.~\ref{fig:Figure2}(b) and \ref{fig:Figure1}(e) confirms that these photovoltage resonances coincide with enhanced optical absorption, establishing a direct link between localized field confinement and photoresponse \cite{Castilla2020, Castilla2024}.

At elevated temperatures ($T=130$~K, Fig.~\ref{fig:Figure2}(c)), the AGP resonances vanish, while the overall six-quadrant PTE symmetry persists. This disappearance is attributed to enhanced electron--phonon scattering, which reduces the carrier mobility $\mu$ and increases the intrinsic damping of collective charge excitations, resulting in a broadened plasmon linewidth $\gamma$ and a reduced plasmon quality factor $Q$~\cite{Koppens2011, Bandurin2018}. $Q$ indeed provides a direct measure of plasmon losses, expressing the balance between energy storage and dissipation during a plasmon oscillation~\cite{Koppens2011, Bandurin2018}. From low-temperature transport characterization (see Supplementary Fig.~S1), we extract mobilities of $\mu_h \approx 3.7\times10^{4}\,\mathrm{cm}^2\,\mathrm{V}^{-1}\,\mathrm{s}^{-1}$ for holes and 
$\mu_e \approx 1.3\times10^{4}\,\mathrm{cm}^2\,\mathrm{V}^{-1}\,\mathrm{s}^{-1}$ for electrons at $T=6$~K in the right-gated region, while $\mu_h \approx 2.7\times10^{4}\,\mathrm{cm}^2\,\mathrm{V}^{-1}\,\mathrm{s}^{-1}$ and 
$\mu_e \approx 1.2\times10^{4}\,\mathrm{cm}^2\,\mathrm{V}^{-1}\,\mathrm{s}^{-1}$ in the left-gated region. We note that the mobility beneath the left gate is lower due to fabrication-induced disorder (see Supplementary Fig.~S1), leading to an asymmetric damping landscape along the channel.

Figures~\ref{fig:Figure2}(d) and (e) further illustrate the temperature dependence of $V_{\mathrm{ph}}$ while sweeping the right-gate voltage $V_{\mathrm{RG}}$ at fixed left-gate voltages $V_{\mathrm{LG}} = -6$~V and $V_{\mathrm{LG}} = 0$~V, respectively. At $V_{\mathrm{LG}} = -6$~V, the Seebeck asymmetry between the left and right gated regions is maximized, resulting in a strong PTE response and pronounced AGP resonances at $V_{\mathrm{RG}} \approx -1$~V and $0.3$~V. As the temperature increases from 6 to 130~K, for example, the amplitude of the first resonance is reduced by at least a factor of $\sim 7$, reflecting the progressive increase of intrinsic plasmon damping with temperature. A more quantitative analysis of the peak-to-background contrast, based on the normalized modulation intensity $M_i(T)$ of each resonance (see Supplementary Fig.~S3), shows that the first AGP mode modulates the photovoltage by $M_1 \approx 40\%$ at $T = 6$~K, decreasing to $\lesssim 15\%$ at $T \approx 130$~K, while the second mode evolves from $M_2 \approx 25$--$30\%$ to below $\sim 5\%$ over the same temperature range. This pronounced reduction of $M_i(T)$ closely follows the expected temperature dependence of $Q$, governed by phonon- and dielectric-limited damping mechanisms~\cite{Ni2018}. While carrier mobility provides a useful qualitative indicator of loss~\cite{Hwang2008}, the lifetime of graphene plasmons is more directly determined by the imaginary part of the plasmon wavevector $q_p$, which can be accessed within our electromagnetic simulations. In the acoustic plasmon regime, the plasmon attenuation obeys $\mathrm{Im}(q_p)=\mathrm{Re}(q_p)/(\omega\tau)$, where $\tau$ is the effective carrier scattering time entering the graphene conductivity, and $\omega = 2\pi f$ is the angular frequency of the plasmon mode. Assuming the approximately linear dispersion relation of AGPs, 
$\omega \approx v_g\,\mathrm{Re}(q_p)\,g$, where $v_g$ is the group velocity of the acoustic plasmon mode and $g=\omega/c$ is the free-space wavenumber, the corresponding plasmon lifetime can be expressed as $\tau_p=[\mathrm{Im}(q_p)\,g\,v_g]^{-1}$, yielding $\tau_p \approx \tau$, and the associated quality factor follows as $Q=\omega\tau_p=\omega\tau$. This framework highlights that the observed AGP resonances are primarily limited by intrinsic graphene losses, with geometric confinement mainly defining the cavity mode structure rather than the damping rate itself.

Applying this analysis to our simulated structures, and since $\tau_p \approx \tau$, we estimate the plasmon lifetime from the carrier scattering time $\tau=\mu|E_F|/v_F^2$ ($v_F=1.15\times10^6~\mathrm{m/s}$), yielding for the second AGP resonance confined under the right gate in the electron-doped regime ($E_F\approx0.13$~eV, $\mu_{e}\approx1.3\times10^{4}\,\mathrm{cm}^2\,\mathrm{V}^{-1}\,\mathrm{s}^{-1}$) a lifetime $\tau_p\approx0.13$~ps and a corresponding quality factor $Q\approx2.0$, whereas for the first resonance, which extends over the full channel and is subject to asymmetric damping (right gate: $E_F\approx0.07$~eV, $\mu_{e}\approx1.3\times10^{4}\,\mathrm{cm}^2\,\mathrm{V}^{-1}\,\mathrm{s}^{-1}$; left gate: $E_F\approx-0.12$~eV, $\mu_{h}\approx2.7\times10^{4}\,\mathrm{cm}^2\,\mathrm{V}^{-1}\,\mathrm{s}^{-1}$), the effective plasmon lifetime obtained by combining the corresponding scattering rates according to $1/\tau_p=1/\tau_{\mathrm{RG}}+1/\tau_{\mathrm{LG}}$ is reduced to $\tau_p\approx0.05$~ps, yielding $Q\approx0.8$. These values are in good agreement with previously reported, relatively low-loss AGP lifetimes in the sub-THz to THz frequency range~\cite{Alonso2017}, and they quantitatively reproduce both the resonance visibility and the temperature-induced suppression observed experimentally in Fig.~\ref{fig:Figure2}. As the temperature increases from 6 to 130~K, the effective reduction of $\tau_p$ inferred from experiment follows the enhanced plasmon damping expected from increased phonon scattering, providing a coherent link between the simulated plasmon losses and the measured temperature dependence.

At $V_{\mathrm{LG}} = 0$~V, where the left-gated region is near its CNP, the Seebeck contrast is minimal, suppressing the overall PTE response and reducing the visibility of the AGP resonances. This confirms that plasmon-enhanced THz detection in our devices relies on the combined effect of (i) strong Seebeck asymmetry to convert absorbed power into measurable photovoltage, and (ii) sufficiently low intrinsic losses in the gated antenna region, for which high carrier mobility provides a useful proxy, to sustain long-lived AGPs. In Fig.~\ref{fig:Figure2}(f) we present a color map of the absolute photovoltage $|V_{\mathrm{ph}}|$ as a function of $V_{\mathrm{RG}}$ and temperature at fixed $V_{\mathrm{LG}} = -6$~V, illustrating the thermal evolution of the AGP resonances. As the temperature increases, both resonant features progressively fade and eventually vanish, consistent with enhanced plasmon damping caused by thermally activated carrier scattering and screening effects~\cite{Principi2013}.

Figure~\ref{fig:Figure3} summarizes the tunability and physical origin of the plasmonic resonances in our split-gated graphene device.  
The experimental data (filled symbols) and theoretical predictions (open symbols) in Fig.~\ref{fig:Figure3}(a) reveal a systematic evolution of the resonance voltages as the left-gate bias $V_{\mathrm{LG}}$ is varied. Two distinct regimes are evident: for $V_{\mathrm{LG}} \lesssim -3~\text{V}$ the resonances follow one trend, whereas for $V_{\mathrm{LG}} \gtrsim 3~\text{V}$ the slopes change noticeably. This crossover occurs close to the charge-neutrality point of the graphene under the left gate ($V_{\mathrm{LG}}^{\mathrm{CNP}} \approx -4~\text{V}$), indicating that the change in doping polarity under the left gate plays an important role in determining the resonance behavior. In addition, both the first (blue) and second (red) resonances gradually shift towards lower $V_{\mathrm{RG}}$ as $V_{\mathrm{LG}}$ increases, although this effect is comparatively weaker. The excellent agreement between experimental measurements and the AGP cavity model over the entire $V_{\mathrm{LG}}$ range validates our interpretation of the device as an electrostatically tunable AGP cavity.

\begin{figure*}[htbp!]
\centerline{\includegraphics[width=180mm]{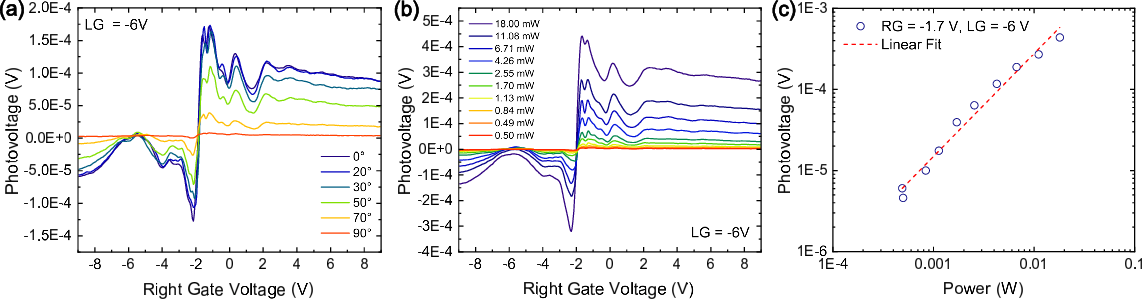}}
\caption{
Power dependence of the device photoresponse at $V_{\mathrm{LG}} = -6$~V.  
(a) Measured photovoltage as a function of right-gate voltage $V_{\mathrm{RG}}$ for different incident powers, controlled using two wire-grid polarizers.  
(b) Independent power dependence obtained by tuning the output power of the laser source, showing a consistent trend over more than one order of magnitude in excitation intensity.  
(c) Log–log plot of the peak photovoltage as a function of incident power, measured at fixed $V_{\mathrm{RG}} = -1.7$~V and $V_{\mathrm{LG}} = -6$~V.  
The nearly linear scaling across almost two decades demonstrates that the device operates in the linear response regime, ensuring predictable and stable performance suitable for THz detection applications.
}
\label{fig:Figure4}
\end{figure*}

To directly relate the measured photovoltage peaks to AGP modes, we compared the experimental data with theoretical predictions based on the AGP dispersion and Fabry--P\'erot conditions~\cite{Alonso2017}. Figs.~\ref{fig:Figure3}(b,c) show the AGP dispersion calculated at the right-gate voltages corresponding to the 1$^{\mathrm{st}}$ and 2$^{\mathrm{nd}}$ resonances observed in the photovoltage spectra (see Fig.~\ref{fig:Figure1}(h)). The colored curves give the plasmon wavelength as a function of frequency $f$, $\lambda_{\mathrm{p}}(f)$, for different left-gate voltages $V_{\mathrm{LG}}$, obtained from the AGP dispersion for a gated graphene sheet~\cite{Voronin2020}:

\begin{equation}
\lambda_{p} = \frac{16\pi \alpha_0 E_F}{g \hbar \omega \left( \varepsilon_{\mathrm{SiO_2}} + \sqrt{ \varepsilon_{\mathrm{SiO_2}}^2 + \dfrac{16 \alpha_0 c E_F \varepsilon_{\mathrm{AlO_x}}}{\hbar \omega^2 d} } \right)},
\label{eq:Lambda}
\end{equation}
where $\alpha_0$ is the fine-structure constant, and $\varepsilon_{\mathrm{AlO_x}} = 9.74$ and $\varepsilon_{\mathrm{SiO_2}} = 4.78$ are the permittivities at $\sim$THz frequencies of the AlO$_x$ layer of thickness $d$ and of the SiO$_2$ substrate, respectively~\cite{Gao2025}.

The horizontal dashed lines in Figs.~\ref{fig:Figure3}(b,c) represent Fabry--P\'erot conditions for two different cavity configurations: a mode spanning the full graphene channel (1$^{\mathrm{st}}$ resonance, Fig.~\ref{fig:Figure3}(b)) and a mode mainly confined beneath the right gate (2$^{\mathrm{nd}}$ resonance, Fig.~\ref{fig:Figure3}(c)). For the full channel of length $L$, the effective plasmon wavelength is given by the harmonic mean of the local wavelengths under the left and right gates~\cite{Lee2019, Castilla2024}:

\begin{equation}
\lambda_{\mathrm{p}}^{\mathrm{eff}} = \frac{2\,\lambda_{\mathrm{p},L}\,\lambda_{\mathrm{p},R}}{\lambda_{\mathrm{p},L}+\lambda_{\mathrm{p},R}} .
\label{eq:HalfChannel}
\end{equation}

The corresponding Fabry--P\'erot resonance condition can be written as:

\begin{equation}
\lambda_{\mathrm{p}}^{\mathrm{eff}} = \frac{2L_{\mathrm{c}}}{m}, \qquad m = 1,2,\dots ,
\label{eq:FP}
\end{equation}

with $L_{\mathrm{c}} = L$ for a mode that extends across the entire channel. For the experimentally observed 1$^{\mathrm{st}}$ resonance, we take $m = 4$, in agreement with the four-node standing-wave pattern in the upper panel of Fig.~\ref{fig:Figure1}(c). Because the graphene channel is electrostatically inhomogeneous, the local plasmon wavelength differs under the two gates, and the total plasmon propagation is determined by both regions, naturally leading to the harmonic-mean form of $\lambda_{\mathrm{p}}^{\mathrm{eff}}$.
When only the right half of the channel of length $L/2$ contributes, the relevant cavity length is $L_{\mathrm{c}} = L/2$ and the local plasmon wavelength under the right gate, $\lambda_{\mathrm{p},R}$, satisfies:

\begin{equation}
\lambda_{\mathrm{p},R} = \frac{2L_{\mathrm{c}}}{m}, \qquad m = 1,2,\dots .
\label{eq:RG}
\end{equation}

The 2$^{\mathrm{nd}}$ resonance is then described by $m = 2$, consistent with the two-node standing-wave pattern observed in the lower panel of Fig.~\ref{fig:Figure1}(c). In both cavity configurations, the intersections of the dispersion curves with the dashed lines yield resonance frequencies close to $2.5~\text{THz}$, in good agreement with the experimental excitation conditions.
A comparison of the field profiles in Fig.~\ref{fig:Figure1}(c) shows that the oscillation pattern under the left gate is very similar for the 1$^{\mathrm{st}}$ and 2$^{\mathrm{nd}}$ resonances, both in shape and in amplitude. This indicates that the lower mobility in the left part of the channel mainly controls the plasmon damping and local field strength in that region for both resonances, but does not by itself determine whether the dominant cavity spans the full channel or only its right half. The transition from a full-channel Fabry--P\'erot mode to a mode localized in the right sub-cavity is instead primarily governed by the carrier-density distribution in the gap, which sets the surface-conductivity profile and the associated AGP reflection at the gap edge (see Fig.~\ref{fig:Figure1}(d) and Supplementary Fig.~S4).
At the 2$^{\mathrm{nd}}$ resonance, the sharp conductivity variation in the gap produces strong plasmon backscattering and, together with the metal contacts, effectively defines a right sub-cavity of length $L_{\mathrm{c}} = L/2$. The Fabry--P\'erot condition is then predominantly sustained by this right sub-cavity, yielding a local plasmon wavelength $\lambda_{\mathrm{p},R} \approx 1.1~\mu\text{m}$, close to $L/2$. By contrast, at the 1$^{\mathrm{st}}$ resonance, the conductivity profile across the gap is much smoother, so AGP reflection at this interface is strongly reduced, and a mode extending across the entire channel is supported. In this case, the effective plasmon wavelength, determined by both halves of the channel, is $\lambda_{\mathrm{p}}^{\mathrm{eff}} \approx 1.18~\mu\text{m}$, close to $L/2 = 1~\mu\text{m}$, as expected for the full-channel Fabry--P\'erot condition. Thus, while the mobility asymmetry sets the level of damping in the left region for both resonances, it is the gap-induced conductivity profile that primarily controls whether the AGP mode behaves as a full-channel cavity or as a right sub-cavity.

\begin{figure*}[htbp!]
\centerline{\includegraphics[width=160mm]{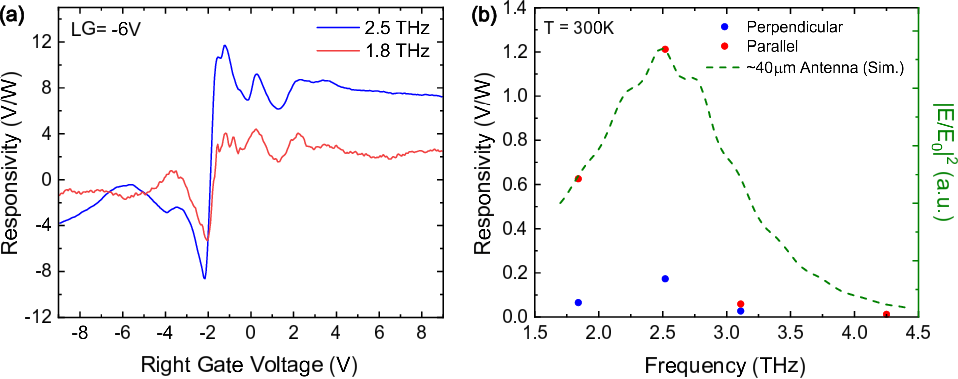}}
\caption{
Frequency dependence of device responsivity. 
(a) Gate-dependent photovoltage measured at T = 6~K, $f = 2.5$~THz (blue) and $f = 1.8$~THz (red), showing reduced response at lower frequency due to weaker antenna coupling.  
(b) Room-temperature peak responsivity as a function of excitation frequency for incident light polarized parallel (red symbols) and perpendicular (blue symbols) to the antenna axis.  
The green dashed line shows simulated normalized near-field enhancement $|E/E_0|^2$ at the graphene plane (right axis), demonstrating excellent agreement between experiment and simulations.  
The responsivity maximum coincides with the designed dipole antenna resonance at $\sim 2.5$~THz, confirming that the device performance is dominated by antenna-mediated coupling.
}
\label{fig:Figure5}
\end{figure*}

To further characterize the device's performance as a THz photodetector, we investigate the power dependence of the measured photovoltage at $V_{\mathrm{LG}} = -6$~V, as shown in Fig.~\ref{fig:Figure4}.  
In Fig.~\ref{fig:Figure4}(a), we control the incident THz power by using a pair of wire-grid polarizers, where one is fixed while the other is rotated to vary the transmitted intensity.  
A systematic reduction of the resonance amplitude is observed as the relative angle between the polarizers increases, resulting in lower incident power on the device.  
The two main resonances remain clearly visible even at strongly attenuated powers, demonstrating the robustness of the photoresponse against variations in optical excitation. An independent power dependence study is reported in Fig.~\ref{fig:Figure4}(b), where the incident THz power is attenuated with THz absorbers. Consistent with the polarizer-based measurements, the photovoltage amplitude scales monotonically with the incident power, confirming the reliability of the response over two orders of magnitude in excitation intensity.

To quantify the detector performance, we fix the gate voltages at $V_{\mathrm{RG}} = -1.7$~V and $V_{\mathrm{LG}} = -6$~V, corresponding to the maximum responsivity, and plot the photovoltage as a function of the incident power in Fig.~\ref{fig:Figure4}(c).  
The resulting log–log plot reveals a nearly linear scaling of the photovoltage over two orders of magnitude of power, with a fitted slope of $\sim$1.  
This linear dependence is a hallmark of a highly efficient and predictable detection regime, confirming that the device operates in the linear response regime across the tested power range.  
Such behavior is particularly promising for THz detection applications, where stable responsivity over a wide dynamic range is a key figure of merit.

\subsection{Frequency Dependence} \label{Freq}

We now investigate the frequency dependence of the device responsivity at $T = 6$~K, as summarized in Fig.~\ref{fig:Figure5}. The device responsivity is calculated as $R_{\mathrm{v}} = (V_{\rm ph}/P_{\rm 0})\times(A_{\rm focus}/A_{\rm diff})$, where $V_{\mathrm{ph}}$ is the measured photovoltage, $P_0$ is the incident THz power, A$_{\rm focus}$ is the experimental beam area at the measured wavelength determined from Gaussian beam diameter measurements, and A$_{\rm diff}$ is the diffraction-limited spot size\cite{Castilla2019, Vicarelli2012, Castilla2020, Viti2021} (see Experimental Section\ref{Sec:Exp} and Supplementary Fig.~S5).
In Fig.~\ref{fig:Figure5}(a), we compare the measured photovoltage response at two different excitation frequencies, $f = 2.5$~THz (blue curve) and $f = 1.8$~THz (red curve), for a fixed $V_{\mathrm{LG}} = -6$~V.   
Both datasets exhibit the same gate-dependent features, corresponding to the acoustic plasmon resonances previously discussed, while the overall responsivity at $1.8$~THz is reduced compared to $2.5$~THz.  
This trend directly reflects the antenna-coupling efficiency, which is expected to peak near the designed dipole resonance at $f\approx 2.5$~THz. The maximum responsivity at 2.5 THz is $\sim$12 V/W, which yields a noise equivalent power (NEP) value of 100 pW/$\sqrt{\rm Hz}$ at 6 K (see \ref{Sec:Exp}Experimental Section).

A more systematic analysis is provided in Fig.~\ref{fig:Figure5}(b), where we plot the room-temperature extracted peak responsivity as a function of frequency for two orthogonal incident polarizations: parallel (red symbols) and perpendicular (blue symbols) to the antenna main axis.  
A clear polarization dependence emerges: the responsivity is maximized when the electric field is aligned with the dipole antenna arms, confirming that efficient coupling occurs when the incident field drives the antenna directly.  
In contrast, when the incident field is polarized perpendicular to the antenna main axis, the photoresponse is strongly suppressed, as the dipolar mode is not properly excited. The overall spectral dependence of the responsivity also agrees well with full-wave simulations of the antenna-coupled device, shown as the green dashed line in Fig.~\ref{fig:Figure5}(b).  
Here, the simulated normalized near-field enhancement $|E/E_0|^2$ at the graphene plane (right axis) closely follows the measured responsivity trend, with a peak around $f \approx 2.5$~THz and a rapid decrease at higher and lower frequencies.  
This consistency confirms that the frequency dependence of the device performance is primarily governed by the antenna resonance, as also supported by the simulated antenna response in Supplementary Fig.~S2.

\section{Conclusions}\label{End}
We have demonstrated a resonant THz photodetector that combines a dipole-antenna split gate with a CVD single-layer graphene channel to harness the PTE and AGP modes. Under 2.5~THz illumination at cryogenic temperature, the device exhibits two robust, gate-tunable resonances that coincide with simulated absorption maxima and standing-wave field patterns. The lower-order mode spans the full channel, whereas the higher-order mode is predominantly confined beneath the right gate, consistent with a Fabry--P\'erot picture in which the carrier-density and conductivity profile along the channel defines the effective cavity length, while mobility asymmetry primarily sets the level of plasmon damping. The photovoltage scales linearly with incident power over nearly two decades, follows the antenna polarization, and peaks at the designed antenna resonance, confirming antenna-mediated coupling as the dominant excitation pathway. As temperature increases, the resonance amplitudes decrease, consistent with enhanced intrinsic plasmon damping arising from thermally activated carrier scattering and screening, for which carrier mobility provides a useful qualitative indicator. At liquid-nitrogen and lower temperatures, the AGP cavity is highly efficient: the lower-order AGP mode alone induces a relative modulation of the PTE photovoltage signal of up to $\sim40\%$, while the higher-order mode produces modulations of up to $\sim25$--$30\%$, demonstrating that the resonant contribution dominates the overall response in the cryogenic regime. Beyond evidencing AGP resonances in potentially scalable, non-encapsulated graphene, our results highlight a practical device architecture in which electrostatic design and antenna engineering jointly define the plasmonic cavity and the thermoelectric readout. Looking forward, higher-quality graphene channels (e.g., hBN-encapsulated CVD graphene), lower-loss dielectrics, and optimized cavity boundaries are expected to reduce intrinsic plasmon losses, increase the quality factor $Q$, and extend operation toward elevated temperatures. Integration with on-chip antennas, impedance-matched readout, and pixelated arrays may enable frequency-selective THz imaging and spectroscopy with low power consumption. More broadly, the platform is compatible with other van der Waals heterostructures and could leverage gain mechanisms (tunneling or superconducting contacts) to approach room-temperature plasmonic THz sensing.

\section{Experimental Section} \label{Sec:Exp}

\textit{Graphene Growth and Transfer}: Monolayer graphene was synthesized by chemical vapor deposition (CVD) on high-purity copper foil (35 $\mu$m-thick, Alfa Aesar) using a 4" Aixtron Black Magic. The growth was carried out under a mixture of methane (CH$_4$) and hydrogen (H$_2$) gases, with flow rates of 30 sccm of CH$_4$, diluted in Ar (0.01\%) and 50 sccm H$_2$, at a growth temperature of $\sim${1000\,$^\circ$C} and pressure of 25\,mbar for 6 hours~\cite{Li2009}. After growth, the samples were rapidly cooled to room temperature under an H$_2$ atmosphere to preserve monolayer quality. The graphene film was transferred onto highly resistive ($\rho>10,000 \Omega\cdot$cm) Si substrates purchased from Siegert Wafer GmbH, with a 300 nm thermally grown SiO$_2$ layer. We used a standard wet-transfer process~\cite{Bae2010}. A thin layer of PMMA (A4-950K, MicroChem) was spin-coated onto the graphene/Cu stack at 4000 rpm for 40 s. The copper foil was subsequently etched away in an aqueous ammonium persulfate (APS) solution ($\sim$3 g in 150 ml of water), followed by multiple rinsing steps in deionized water to remove etchant residues. The graphene/PMMA membrane was carefully scooped and transferred onto the target substrate, followed by drying at room temperature for 8 hours. Finally, PMMA was removed by immersion in acetone for 1 hour, followed by a rinse in isopropanol and blow-drying in nitrogen, which leaves monolayer graphene on the target substrate. To verify the quality of the transferred graphene, Raman spectroscopy was performed using a Renishaw inVia confocal Raman microscope equipped with a 532 nm excitation laser and a 100$\times$ objective lens, as reported in the Supplementary Material, Fig.~S6. The Raman analysis confirmed the preservation of high-quality monolayer graphene before and after transfer. 

\textit{Device Fabrication}: Source and drain contacts were patterned using a maskless aligner lithography system (using Heidelberg MLA150, from Heidelberg Instruments Mikrotechnik GmbH). The photoresist used was ECI 3007 (positive tone), which was spin-coated at 4000 rpm for 60 seconds, followed by a soft bake at 100\,$^\circ$C for 1 minute. After exposure, a post-exposure bake (PEB) was performed at 110\,$^\circ$C for 1 minute before development, followed by thermal evaporation of Ti (5 nm) / Au (25 nm) and lift-off in acetone. The resulting source-drain separation was 2 $\mu$m, matching the designed channel width. The graphene channel was subsequently patterned into an H-shape via reactive ion etching (RIE) using an O$_2$ plasma, RF power of 3 W, and etch time of 30 seconds, a geometry chosen to optimize the contact resistance between graphene and the metal electrodes. A $\sim40$ nm-thick AlO$_x$ layer was then deposited by atomic layer deposition (ALD) using trimethylaluminium (TMA) and water as precursors, at a deposition temperature of 250\,$^\circ$C, forming a uniform high-$\kappa$ gate dielectric. Finally, the split-gate dipole antenna structure was patterned via electron-beam lithography (EBL) with a PMMA 950 K resist film, followed by electron-beam evaporation of Ti (5 nm) / Au (50 nm) and lift-off in acetone. The resulting two antenna lobes act as both efficient terahertz (THz) coupling elements and independent electrostatic gates, separated by a 200 nm gap defined by the EBL resolution limit. The total antenna length was designed to be 40 $\mu$m, optimized through full-wave electromagnetic simulations (see Supplementary Material, Fig.~S2).

\textit{Optoelectronic Setup}: We use a continuous wave THz beam from a gas laser (FIRL 100 from Edinburgh Instruments), delivering maximum output powers in the range of a few tens of milliwatts and at discrete emission lines of 1.83, 2.52, 3.11, and 4.25 THz. The THz laser was modulated at 267 Hz using an optical chopper, and the resulting photovoltage was measured using a lock-in amplifier (Stanford Research Systems). Polarization analysis confirmed that the emitted radiation was strongly linearly polarized, with less than 2$\%$ residual unpolarized contribution. The detector was oriented such that the antenna axis was aligned with the polarization axis. The THz beam was focused into the cryostat via a 20 mm focal length lens mounted on a cage system, and the incident power was quantified using a pyroelectric detector (Gentec-EO) positioned in the optical path.

The sample was mounted in a closed-cycle cryostat (Advanced Research Systems, 4 K base temperature) equipped with optical access and active temperature control. Polarization rotation was achieved using wire-grid polarizers (Purewave) in combination with crystalline quartz quarter-wave plates (Tydex). In this configuration, the quarter-wave plate rotated the polarization state of the incident beam, while the subsequent polarizer allowed for arbitrary selection of the polarization direction. The photoresponse was verified to scale linearly with incident power.

\textit{Responsivity and NEP calculation}: The external responsivity $R_{\rm v}$ is given by~\cite{Castilla2019, Castilla2020, Viti2021}:

\begin{equation}
R_{\rm v} = \frac{V_{\rm ph}}{P_{\rm eff}} = \frac{V_{\rm ph}}{P_{0}\,(A_{\rm diff}/A_{\rm focus})},
\label{eq:responsivity}
\end{equation}

Where $P_{0}$ is the power measured by the commercial power meter, $A_{\rm focus}$ is the experimentally measured beam area at the sample plane, and $A_{\rm diff}$ is the diffraction-limited spot size. The photovoltage $V_{\rm ph}$ is obtained from the output of the lock-in amplifier ($V_{\rm LIA}$) using $V_{\rm ph} = \frac{2\pi\sqrt{2}}{4} V_{\rm LIA}$~\cite{Castilla2019, Viti2021}.
Since the experimental beam is much larger than a diffraction-limited spot, only the fraction $A_{\rm diff}/A_{\rm focus}$ of the measured power $P_{0}$ can, in principle, be concentrated onto the detector. This defines an effective power $P_{\rm eff} = P_{0}(A_{\rm diff}/A_{\rm focus})$. In our experiments, this fraction is $A_{\rm diff}/A_{\rm focus} \approx 1/240 \approx 4\times10^{-3}$. The ratio between the diffraction-limited and experimental areas is:

\begin{equation}
\frac{A_{\rm diff}}{A_{\rm focus}} =
\frac{w_{0,\rm diff}^{2}}{w_{0,x}\,w_{0,y}},
\label{eq:arearatio}
\end{equation}

Where the beam radii $w_{0,x}$ and $w_{0,y}$ are extracted by scanning the device in the $x$- and $y$-directions and fitting the photocurrent profiles with Gaussian functions $\propto e^{-2x^{2}/w_{0,x}^{2}}$ and $\propto e^{-2y^{2}/w_{0,y}^{2}}$. These radii are related to the standard deviation via $\sigma = w_{0}/2$ and to the full width at half maximum (FWHM) via $\mathrm{FWHM} = \sqrt{2\ln 2}\,w_{0}$. At $\lambda = 120~\mu$m we typically obtain $w_{0,x} \approx w_{0,y} = 594~\mu$m (see Fig.~S3). For the diffraction-limited spot we use $w_{0,\rm diff} = \lambda/\pi$, giving $A_{\rm diff} = \pi w_{0,\rm diff}^{2} = \lambda^{2}/\pi$. Finally, the noise-equivalent power (NEP) is defined as $\mathrm{NEP} = V_{\rm noise}/R_{\rm v}$ and, since the unbiased device exhibits very low intrinsic noise dominated by Johnson noise, we use a noise spectral density $V_{\rm noise} = \sqrt{4k_{\rm B}TR_{\rm D}}$, where $k_{\rm B}$ is the Boltzmann constant, $T$ is the operating temperature, and $R_{\rm D}$ the device resistance.

\section{Acknowledgements} \label{Ackn}
This work has been funded by the PRIN PNRR 2022 project "Continuous THERmal monitoring with wearable mid-InfraRed sensors (THERmIR)" (code P2022AHXE5, CUP 53D23007320001), by the INTERREG VI-A Italy–Croatia 2021–2027 project titled “Civil Protection Plan Digitalization through Internet of Things Decision Support System based Platform (DIGITAL PLAN)” (code ITHR020043, CUP H75E23000200005), and by the PRIMA 2023 project "Food value chain intelligence and integrative design for the development and implementation of innovative food packaging according to bioeconomic sustainability criteria (QuiPack)" (CUP H73C23001270005). TS and LMM acknowledge projects PID2023-148359NB-C21 and CEX2023-001286-S (financed by MICIU/AEI/10.13039/501100011033) and the Government of Aragón through Project Q-MAD. A.P. acknowledges support from the Leverhulme Trust under the grant agreement RPG-2023-253. Furthermore, the research leading to these results has received funding from the European Union Seventh Framework Programme under grant agreement no.785219 and no. 881603 Graphene Flagship for Core2 and Core3. S.C., S.M., K.P.S., and F.H.L.K. acknowledge PDC2022-133844-I00, funded by MCIN/AEI/10.13039/501100011033 and by the “European Union NextGenerationEU/PRTR". S.C., K.P.S., and F.H.L.K. acknowledge funding for the ERC PoC project TERACOMM, grant no. 101113529. Views and opinions expressed are, however, those of the author(s) only and do not necessarily reflect those of the European Union or the European Research Council Executive Agency. Neither the European Union nor the granting authority can be held responsible for them.

\bibliography{literature_merged}

\clearpage


\maketitle

\renewcommand{\thefigure}{S\arabic{figure}}
\renewcommand{\thetable}{S\arabic{table}}
\renewcommand{\theequation}{S\arabic{equation}}
\renewcommand{\thepage}{S\arabic{page}}
\setcounter{figure}{0}
\setcounter{table}{0}
\setcounter{equation}{0}
\setcounter{page}{1}


\section{Supplementary Materials to: Enhanced Terahertz Photoresponse via Acoustic Plasmon Resonances in Graphene}

\section{Electrical Mobility of Graphene}

\begin{figure*}[htbp!]
\centerline{\includegraphics[width=180mm]{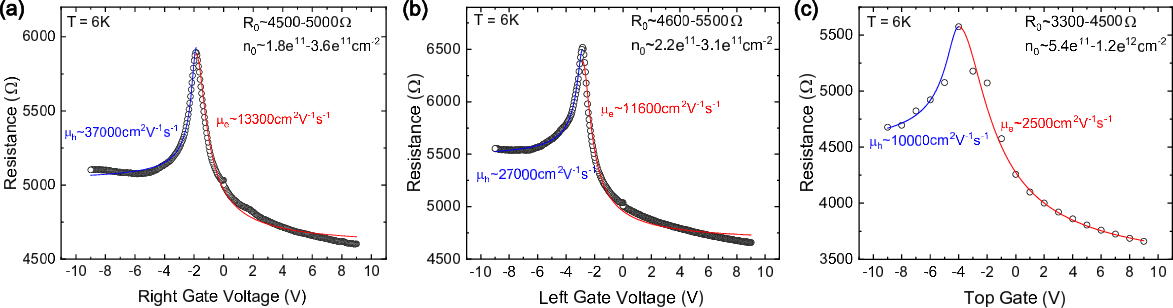}}
\caption{Low-temperature ($T=6$~K) transport characterization of the device. 
(a) Two-terminal resistance vs.\ right-gate voltage, yielding $\mu_h \approx 3.7\times10^4$~cm$^2$V$^{-1}$s$^{-1}$ and $\mu_e \approx 1.33\times10^4$~cm$^2$V$^{-1}$s$^{-1}$.
(b) Left-gate sweep, yielding $\mu_h \approx 2.7\times10^4$~cm$^2$V$^{-1}$s$^{-1}$ and $\mu_e \approx 1.16\times10^4$~cm$^2$V$^{-1}$s$^{-1}$.
(c) Uniform gating configuration, showing reduced mobilities $\mu_h \approx 1.0\times10^4$~cm$^2$V$^{-1}$s$^{-1}$ and $\mu_e \approx 2.5\times10^3$~cm$^2$V$^{-1}$s$^{-1}$. 
Extracted $R_0$ and $n_0$ values are indicated in each panel.}
\label{fig:FigureS1}
\end{figure*}

Supplementary Fig.~\ref{fig:FigureS1} shows the electrical characterization of the graphene channel used in this work at $T=6$~K. The experimental data are fitted using the following model:

\begin{equation}
R = R_0 + \frac{L}{W} \cdot \frac{1}{e \mu \sqrt{n_0^2 + \left( \frac{C_{\mathrm{ox}}(V - V_{\mathrm{CNP}})}{e} \right)^2 }},
\end{equation}

where $R_0$ is the residual contact resistance, $L=W=2~\mu\mathrm{m}$, and 
$C_{\mathrm{ox}}=\varepsilon_0\varepsilon_r/t_{\mathrm{ox}}$ 
is the gate capacitance per unit area.  
In this work, gating is achieved through the top AlO$_x$ dielectric layer of thickness $t_{\mathrm{ox}} = 40~\mathrm{nm}$, 
with $\varepsilon_0 = 8.85\times 10^{-12}~\mathrm{F\,m^{-1}}$ the vacuum permittivity.  
Using a (DC) dielectric permittivity of $\varepsilon_r \approx 5$, for ALD-grown AlO$_x$ films~\cite{Kolodzey2000},  we obtain $C_{\mathrm{ox}} = 9.96\times10^{-8}~\mathrm{F\,cm^{-2}}$. This $C_{\mathrm{ox}}$ value is used consistently for all mobility extractions.

Panel (a) displays the resistance $R$ as a function of right-gate voltage $V_{\mathrm{RG}}$ with the left gate grounded, 
yielding mobilities of $\mu_e \approx 1.33\times10^{4}\,\mathrm{cm}^2\mathrm{V^{-1}s^{-1}}$ for electrons and 
$\mu_h \approx 3.7\times10^{4}\,\mathrm{cm}^2\mathrm{V^{-1}s^{-1}}$ for holes.  
Panel (b) reports the corresponding data for the left-gated region ($V_{\mathrm{LG}}$ sweep), where we extract 
$\mu_e \approx 1.16\times10^{4}\,\mathrm{cm}^2\mathrm{V^{-1}s^{-1}}$ and $\mu_h \approx 2.7\times10^{4}\,\mathrm{cm}^2\mathrm{V^{-1}s^{-1}}$.  
Panel (c) shows the resistance when sweeping both gates simultaneously, resulting in symmetric doping across the entire channel; here, 
the extracted electron mobility is $\mu_e \approx 2.5\times10^{3}\,\mathrm{cm}^2\mathrm{V^{-1}s^{-1}}$, while hole mobility is $\mu_h \approx 1.0\times10^{4}\,\mathrm{cm}^2\mathrm{V^{-1}s^{-1}}$, indicating stronger scattering in the central graphene region. The significant difference in mobility between the left and right gate regions is consistent with additional disorder introduced during fabrication, as discussed in the main text.

Furthermore, the fitting procedure also provides estimates for the contact resistance $R_0$ and the residual carrier density $n_0$, which are key figures of merit for the device quality.  
From Figs.~\ref{fig:FigureS1}(a-c), we extract $R_0 \sim 3.3{-}5.5~\mathrm{k\Omega}$ for the three configurations, consistent with low-resistance Ti/Au contacts optimized through the H-shaped graphene geometry. The extracted residual carrier densities are $n_0 \sim 1.8{-}3.6\times10^{11}~\mathrm{cm^{-2}}$ for the right-gated configuration, $n_0 \sim 2.2{-}3.1\times10^{11}~\mathrm{cm^{-2}}$ for the left-gated case, and up to $n_0 \sim 5.4{-}12\times10^{11}~\mathrm{cm^{-2}}$ when both gates are swept simultaneously. The larger $n_0$ observed in the uniformly doped case reflects stronger disorder and trapped charges in the channel, which partly explains the degraded mobility.

\section{Antenna Optimization}

 \begin{figure}[htbp!]
\centerline{\includegraphics[width=80mm]{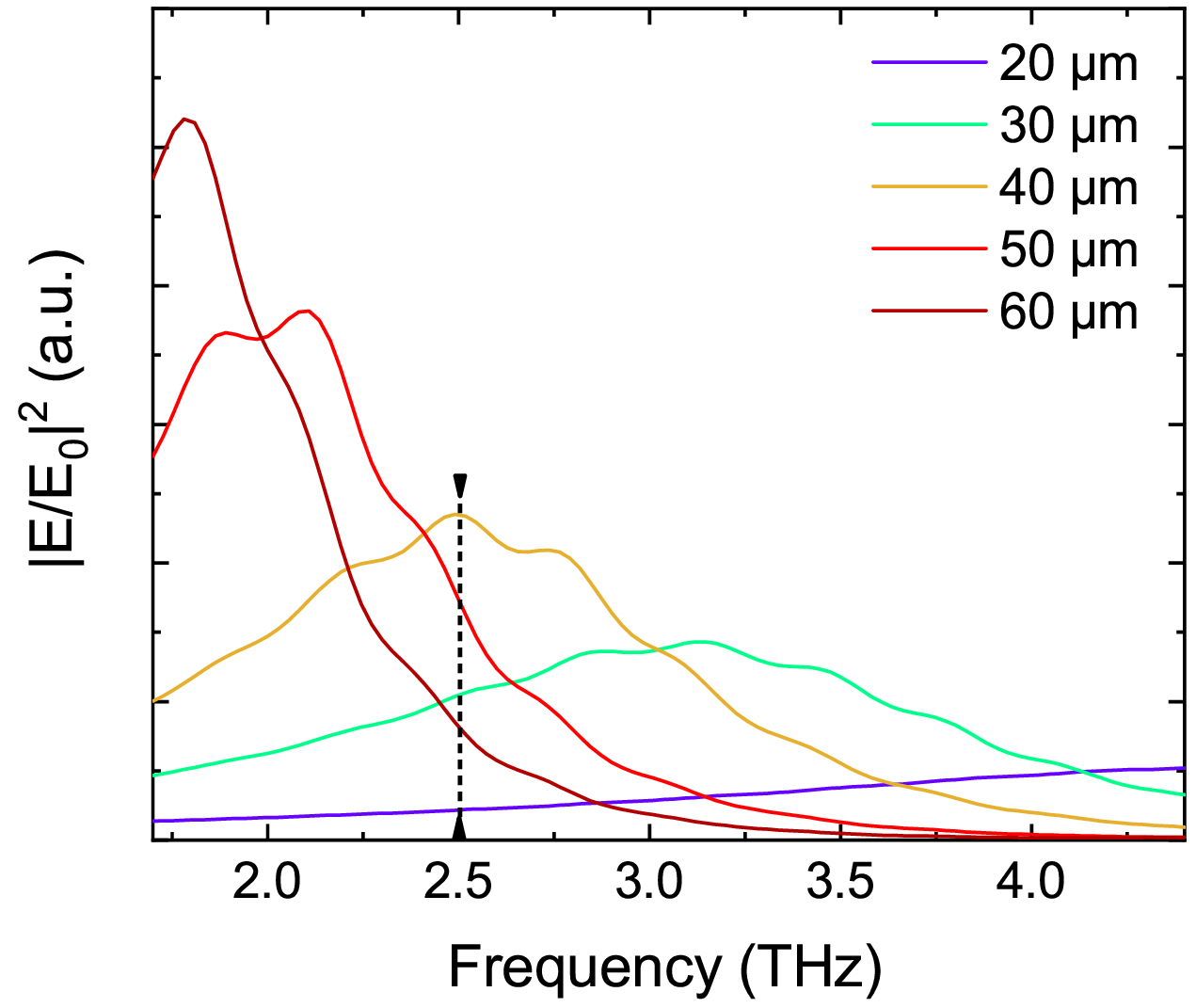}}
\caption{
Field intensity as a function of the incident THz frequency for different dipole antenna lengths. The length includes both branches (left and right) of the antenna and the gap between them. The black arrow indicates the targeted resonance at 2.5 THz, corresponding to the fabricated antenna length of 40 \textmu m.
}
\label{fig:FigureS2}
\end{figure}

The device is optimized by performing optical calculations with the full vector 3D finite-difference time domain (FDTD) method using Lumerical software. We use perfectly matched layer (PML) conditions employed on all boundaries of the computational cell. More information on the photoresponse modelling is described in \cite{Castilla2020, Vangelidis2022}. The stack is made up of two 30 nm thick nanorods separated by a 200 nm gap (the half-wave dipole antenna), a 40 nm thick alumina layer, a graphene layer, a 2 $\mu$m thick SiO$_2$ layer, and a Silicon substrate. The graphene layer is modeled using an H-shaped 2D conductive material having a 2 $\mu$m wide central region. The optical properties of the dielectric layers are described by means of a complex permittivity model, while for graphene, a complex conductivity model is used. The excitation source is a Total Field Scattered Field (TFSF) source. 
The length of the nanorods is swept in the range 20 $\mu$m - 60 $\mu$m to tune the resonance frequency at the desired value of 2.5 THz. As it is possible to observe in Fig.~\ref{fig:FigureS2}, where the electric field enhancement at the center of the antenna's gap is reported, such a value of resonance frequency is obtained for a length of the nanorods equal to 40 $\mu$m. 

\section{Temperature–Dependent Modulation Intensity of the AGP Resonances}

\begin{figure}[htbp!]
\centerline{\includegraphics[width=85mm]{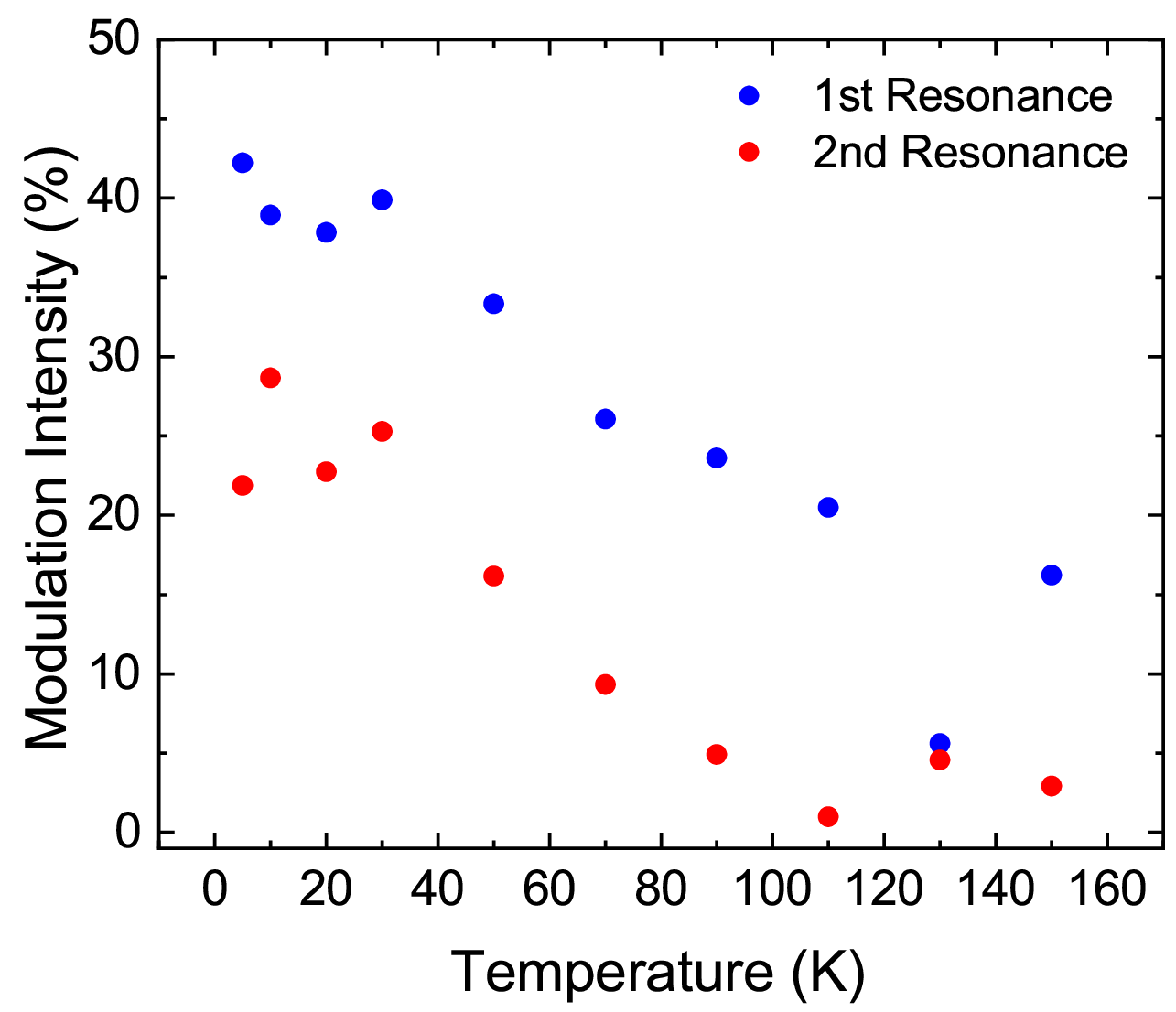}}
\caption{Modulation intensity $M_i(T)$ of the first (blue symbols) and second (red symbols) AGP resonances, extracted from the right-gate sweeps in Fig.~2(d) of the main text at $V_{\mathrm{LG}} = -6~\mathrm{V}$ and $f = 2.5~\mathrm{THz}$. The modulation intensity is defined as the normalized peak-to-trough contrast between the resonance maximum and the subsequent minimum encountered while sweeping $V_{\mathrm{RG}}$ in the same direction [Eq.~(\ref{eq:modulation_intensity})].}
\label{fig:FigureS3}
\end{figure}

To quantify how strongly the AGP resonances modulate the PTE response, we introduce a temperature-dependent \emph{modulation intensity} for each resonance. This analysis is based on the right-gate sweeps $V_{\mathrm{RG}}$ at fixed $V_{\mathrm{LG}} = -6~\mathrm{V}$ under $f = 2.5~\mathrm{THz}$ illumination and for temperatures between $T = 6$ and $150~\mathrm{K}$, \textit{i.e.}, on the same dataset used in Fig.~2(d,f) of the main text.

For each temperature $T$, we consider the absolute photovoltage $|V_{\mathrm{ph}}(V_{\mathrm{RG}},T)|$ and identify the local maximum $|V_{\mathrm{ph}}^{\max,i}(T)|$ associated with the $i$-th AGP resonance ($i=1$ for the 1$^{\mathrm{st}}$ resonance and $i=2$ for the 2$^{\mathrm{nd}}$ resonance). Starting from this maximum and continuing the sweep in the same direction of increasing $V_{\mathrm{RG}}$, we then locate the nearest local minimum
$|V_{\mathrm{ph}}^{\min,i}(T)|$ that represents the background response after the system has been driven through the corresponding resonance. We define the modulation intensity of the $i$-th resonance at temperature $T$ as the normalized peak-to-trough contrast:

\begin{equation}
  M_i(T) = 
  \frac{|V_{\mathrm{ph}}^{\max,i}(T)| - |V_{\mathrm{ph}}^{\min,i}(T)|}
       {|V_{\mathrm{ph}}^{\max,i}(T)|} \, ,
  \label{eq:modulation_intensity}
\end{equation}

and express $M_i(T)$ in percent. By construction, $M_i(T)=0$ corresponds to a completely unmodulated response (no discernible resonance), whereas $M_i(T)\rightarrow 100\%$ would indicate a fully quenched background between the peak and the nearby minimum. This definition is similar to the contrast metrics commonly used in near-field measurements of graphene plasmons, where the relative modulation of the signal encodes the plasmon quality factor and propagation length~\cite{Ni2018}.

The extracted modulation intensities for the first and second resonances are summarized in Supplementary Fig.~\ref{fig:FigureS3}. At the lowest temperatures ($T \lesssim 20~\mathrm{K}$), the first resonance exhibits a modulation intensity $M_1$ of approximately $40$--$45\%$, while the second resonance reaches $M_2 \approx 25$--$30\%$. As the temperature is increased, both resonances show a monotonic reduction of $M_i(T)$: around $T \approx 90~\mathrm{K}$ we find $M_1 \approx 25\%$ and $M_2 \approx 10$--$15\%$, whereas at $T \approx 130$--$150~\mathrm{K}$ the modulation intensities fall below $\sim 15\%$ and $\sim 5\%$ for the first and second resonance, respectively. This quantitative trend confirms that the AGP resonances provide a substantial modulation of the PTE response at low temperatures and progressively fade as plasmon damping increases. The systematically larger modulation intensity of the first resonance can be understood within the cavity picture developed in the main text. The first mode corresponds predominantly to an AGP standing wave that spans the full channel, with an electric-field maximum near the antenna gap and strong overlap with the region of largest Seebeck asymmetry. In contrast, the second mode is more localized toward the right half of the channel and is therefore more susceptible to intrinsic plasmon losses arising from disorder and spatial variations of the transport properties along the channel, which effectively reduce the cavity quality factor and hence the modulation depth. This behaviour is fully consistent with the overall temperature evolution of the AGP resonances discussed in the main text, where a reduction of carrier mobility by nearly two orders of magnitude between $T = 6$ and $130~\mathrm{K}$ is accompanied by enhanced plasmon damping and the eventual disappearance of the resonant features. More broadly, the temperature dependence of $M_i(T)$ mirrors the trend reported for the plasmonic quality factor $Q = Q(T)$ in cryogenic near-field imaging experiments on high-mobility encapsulated graphene~\cite{Ni2018}. In those studies, a rapid increase of $Q$ and of the plasmon propagation length is observed upon cooling, reflecting a dramatic reduction in the intrinsic plasmonic scattering rate. Our modulation-intensity analysis shows that, even in a CVD-grown and non-encapsulated graphene platform, the AGP cavity retains a strong resonant character at liquid-nitrogen and lower temperatures: the first AGP mode alone can modulate the PTE photoresponse by $\sim 40\%$ at $T = 6~\mathrm{K}$, highlighting the potential of such structures for plasmonically enhanced THz detection.

\section{Conducibility}

\begin{figure}[htbp!]
\centerline{\includegraphics[width=85mm]{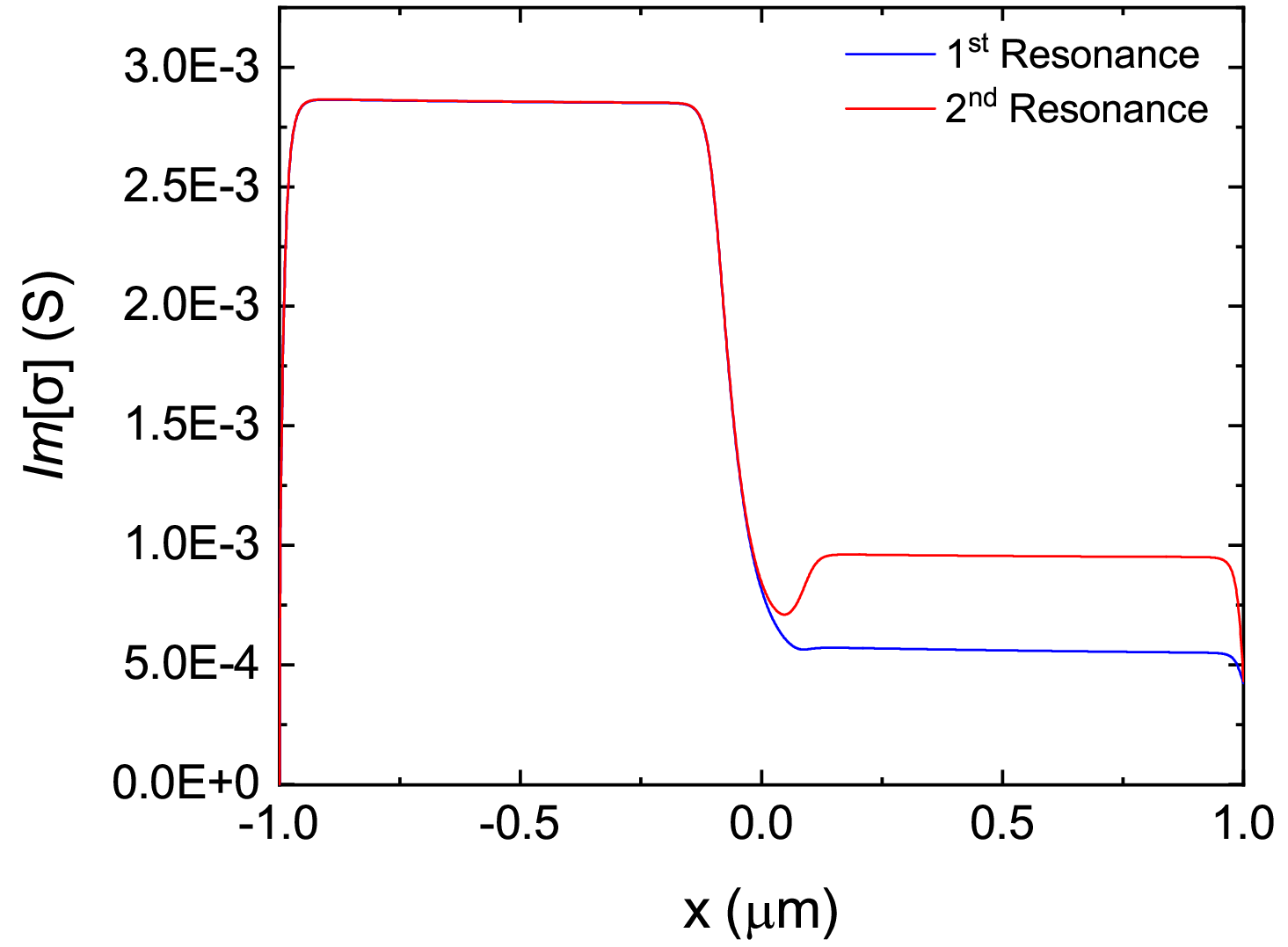}}
\caption{Imaginary part of the graphene conductivity in the case of the 1$^{\mathrm{st}}$ and 2$^{\mathrm{nd}}$ resonance.}
\label{fig:FigureS4}
\end{figure}

The reason why, in the case of the 1$^{\mathrm{st}}$ resonance, the entire graphene channel participates in the Fabry---Perot mode, whereas for the 2$^{\mathrm{nd}}$ resonance, only half of the channel is effectively involved, is non-trivial. Here, we attribute it to the combined effect of graphene mobility asymmetry (under the two gates) and an abrupt change in surface conductivity in the gap (see Fig.~\ref{fig:FigureS4}). The lower mobility under the left gate increases plasmon damping and hinders efficient plasmon propagation in that region, while the abrupt conductivity change in the gap region acts as an additional plasmon reflector. As a result, in the 2$^{\mathrm{nd}}$ resonance case, the Fabry--Perot condition is primarily sustained by the higher-mobility right side, yielding a local plasmon wavelength of $\lambda_\mathrm{p}^{(2)}$ close to the half-channel resonance. By contrast, at the 1$^{\mathrm{st}}$ resonance, the effective plasmon wavelength is determined by both halves of the channel, $\lambda_\mathrm{p}^{(1)}$, corresponding to the full-channel Fabry--Perot condition. In this case, the smoother conductivity profile across the channel suppresses plasmon backscattering at the gap region, enabling a standing wave that extends across the entire device despite the higher losses on the left side.

\section{Beam Diameter}

To determine the effective beam diameter at the device plane, the THz spot was scanned laterally across the detector using a motorized servo stage. At each position $x$, the photocurrent was recorded at a fixed frequency of $f = 2.5$ THz. The resulting spatial profile, shown in Supplementary Fig.~\ref{fig:FigureS5}, exhibits a Gaussian shape centered on the device location. Fitting the data with a Gaussian function:

\begin{equation}
I_{\mathrm{ph}}(x) = I_0 \exp\!\left[-\frac{(x - x_0)^2}{2\sigma^2}\right],
\end{equation}

yields a full width at half maximum (FWHM) of $\mathrm{FWHM} = 2\sqrt{2\ln{2}}\,\sigma \approx 0.7~\mathrm{mm}$. This corresponds to a $1/e^2$ beam radius $w_0 = \mathrm{FWHM}/\sqrt{2\ln 2} \approx 0.59~\mathrm{mm}$, consistent with the values $w_{0,x} \approx w_{0,y}$ used in the Experimental Section.
The illuminated beam area at the sample is then $A_{\mathrm{focus}} = \pi w_{0,x} w_{0,y}$, which we identify with the experimental beam area $A_{\mathrm{focus}}$ entering the responsivity definition $R_{\rm v} = {V_{\rm ph}}/{P_{\rm eff}} = {V_{\rm ph}}/[{P_{0}\,(A_{\rm diff}/A_{\rm focus})}]$ in the main text.

\begin{figure}[htbp!]
\centerline{\includegraphics[width=85mm]{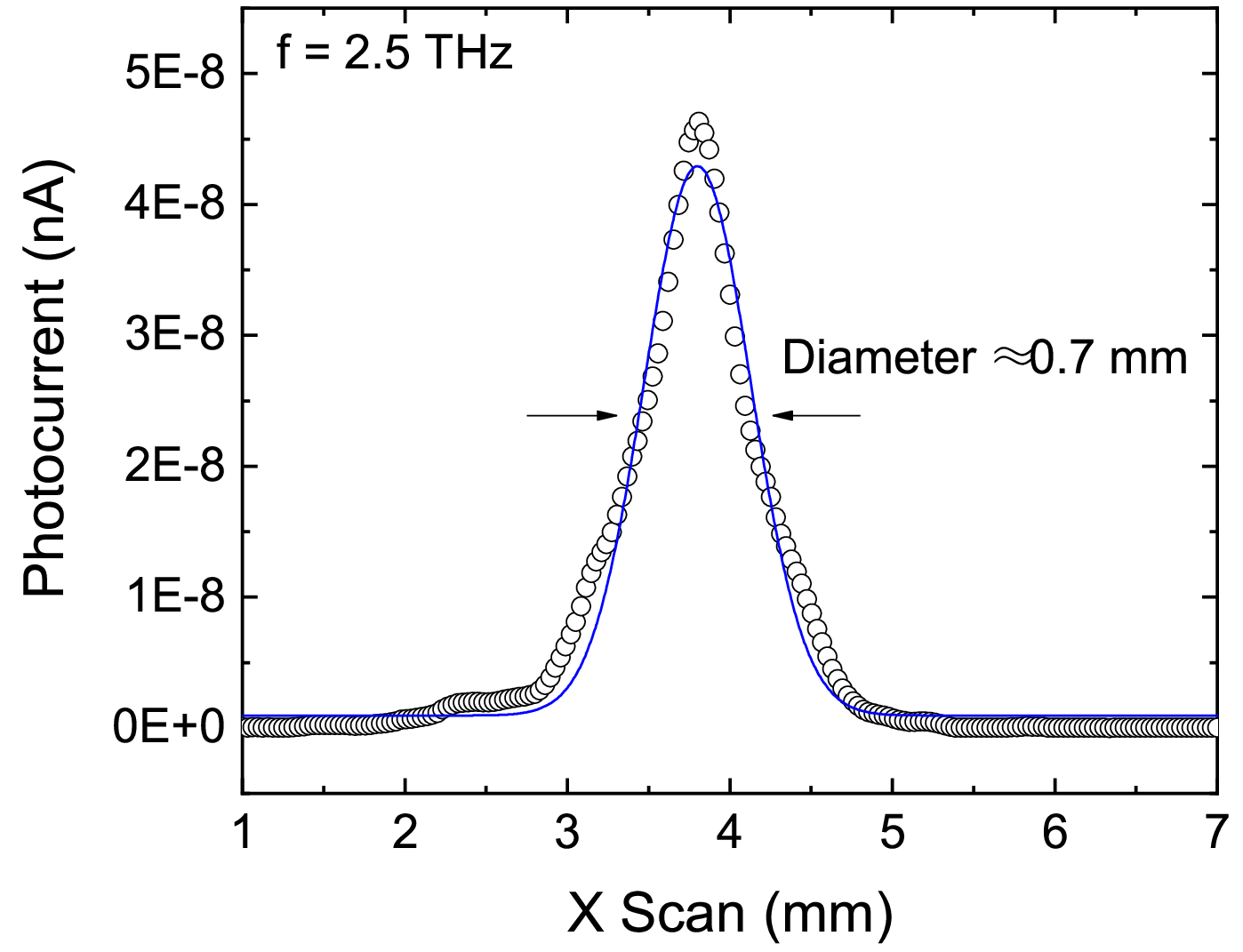}}
\caption{
Determination of the THz beam diameter at the sample position. The photocurrent is measured as a function of lateral beam position ($x$-scan) at $f = 2.5$~THz. Open symbols represent the experimental data, while the solid line is a Gaussian fit.}
\label{fig:FigureS5}
\end{figure}

\section{Raman Spectroscopy of Graphene}

\begin{figure}[htbp!]
\centerline{\includegraphics[width=85mm]{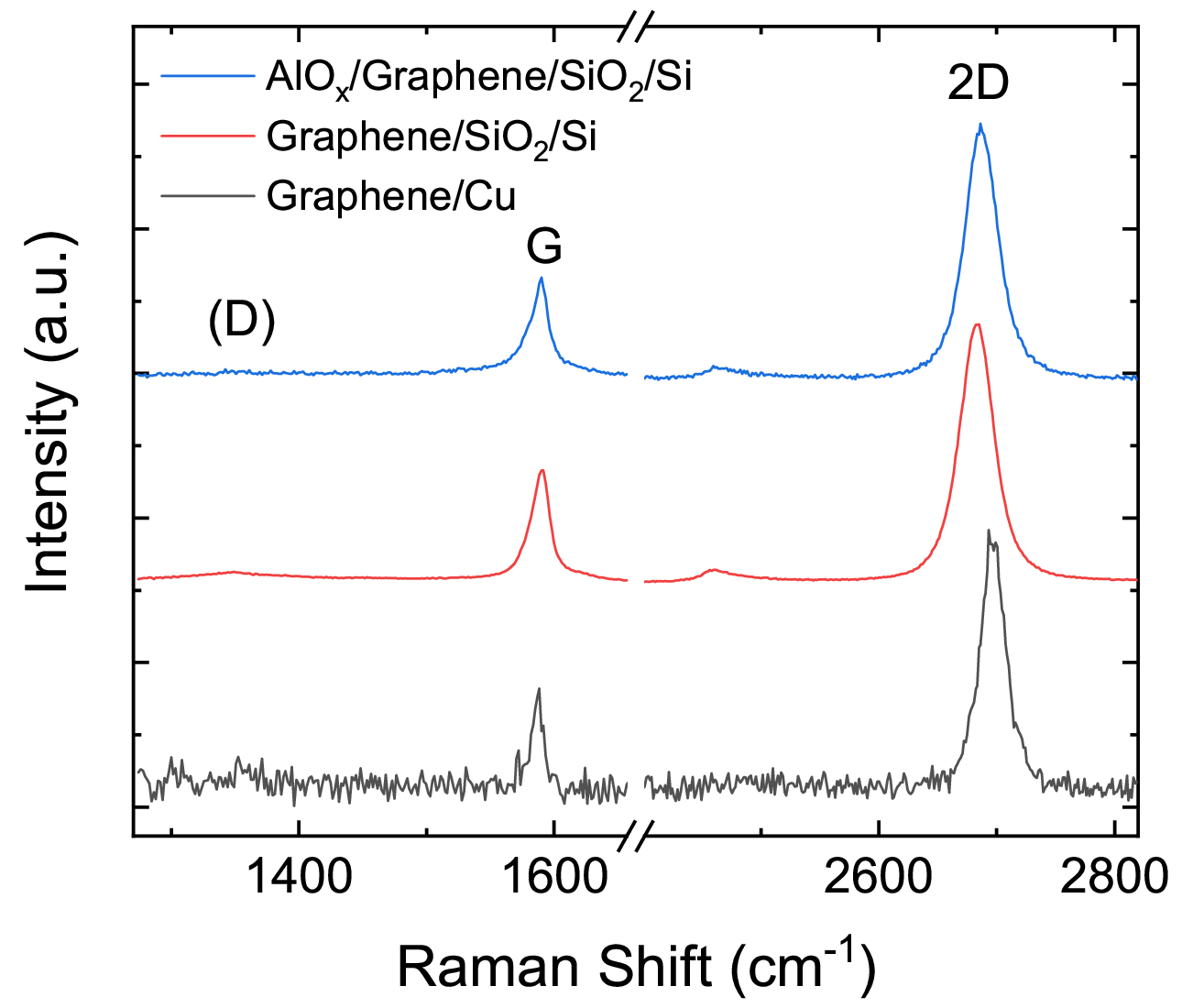}}
\caption{
Raman spectra of graphene at different stages of device fabrication: as-grown on Cu (black), after wet transfer onto SiO$_2$/Si (red), and after AlO$_x$ deposition (blue).  
All spectra are normalized to the maximum peak intensity and vertically offset for clarity.  
Measurements were performed using a Renishaw InVia spectrometer with a 532~nm excitation wavelength and 100$\times$ objective, maintaining the incident laser power below 1~mW to prevent sample damage.}
\label{fig:FigureS6}
\end{figure}

Supplementary Fig.~\ref{fig:FigureS6} shows the Raman spectra of monolayer graphene acquired at different stages of the fabrication process.  
All measurements were taken with a Renishaw InVia Raman spectrometer equipped with a 100$\times$ objective lens and an excitation wavelength of 532~nm, while keeping the laser power below 1~mW to avoid sample heating or degradation.  
The spectra were normalized to the height of the maximum peak, and for the as-grown graphene on Cu, the broad photoluminescence background from the metallic substrate was subtracted for clarity~\cite{Lagatsky2013}.

The bottom (black) curve corresponds to graphene directly after CVD growth on Cu foil.  
The middle (red) curve shows the spectrum after wet transfer onto SiO$_2$/Si, while the top (blue) curve corresponds to the graphene after AlO$_x$ encapsulation.  For graphene on SiO$_2$/Si, the 2D peak appears at $\text{Pos(2D)}\sim$ 2683~cm$^{-1}$ with a full width at half maximum $\text{FWHM(2D)}\sim$ 33~cm$^{-1}$. The G peak is located at $\text{Pos(G)}\sim$ 1590~cm$^{-1}$ with $\text{FWHM(G)}\sim$ 17~cm$^{-1}$. The area ratio $\text{A(2D)/A(G)}\sim$ 4.6, while the intensity ratio $\text{I(2D)/I(G)}\sim$2.4.  
These values are consistent with monolayer graphene exhibiting a single-Lorentzian 2D peak profile and a $\sim$200-250 meV doping~\cite{Ferrari2006, Ferrari2013, Das2008, Basko2009}. Upon AlO$_x$ deposition, the Raman peaks remain sharp and well-defined.  
The G peak appears at $\text{Pos(G)}\sim$ 1589~cm$^{-1}$ with $\text{FWHM(G)}\sim$ 18~cm$^{-1}$, while the 2D peak is centered at $\text{Pos(2D)}\sim$ 2687~cm$^{-1}$ with $\text{FWHM(2D)}\sim$ 32~cm$^{-1}$. The intensity and area ratios are $\text{I(2D)/I(G)}\sim$ 2.6 and $\text{A(2D)/A(G)}\sim$ 5.1, respectively. These values are in line with a $\sim$200-250 n-type doping, corroborating our electrical measurements~\cite{Ferrari2006, Ferrari2013, Das2008, Basko2009}. Besides, they indicate that the AlO$_x$ deposition induces only minor variations, without any evidence of structural degradation or defect formation. No measurable D peak at $\sim1350$~cm$^{-1}$ is observed in any of the spectra, indicating that neither the transfer nor the dielectric deposition introduces additional defects within the detection limit~\cite{Cancado2011, Basko2009, Das2008, Ferrari2006}.  Overall, the Raman analysis confirms that the graphene retains its monolayer character and high crystalline quality throughout the full device fabrication sequence.

\section{Photovoltage as a Function of Left-Gate Voltage}

\begin{figure}[htbp!]
\centerline{\includegraphics[width=85mm]{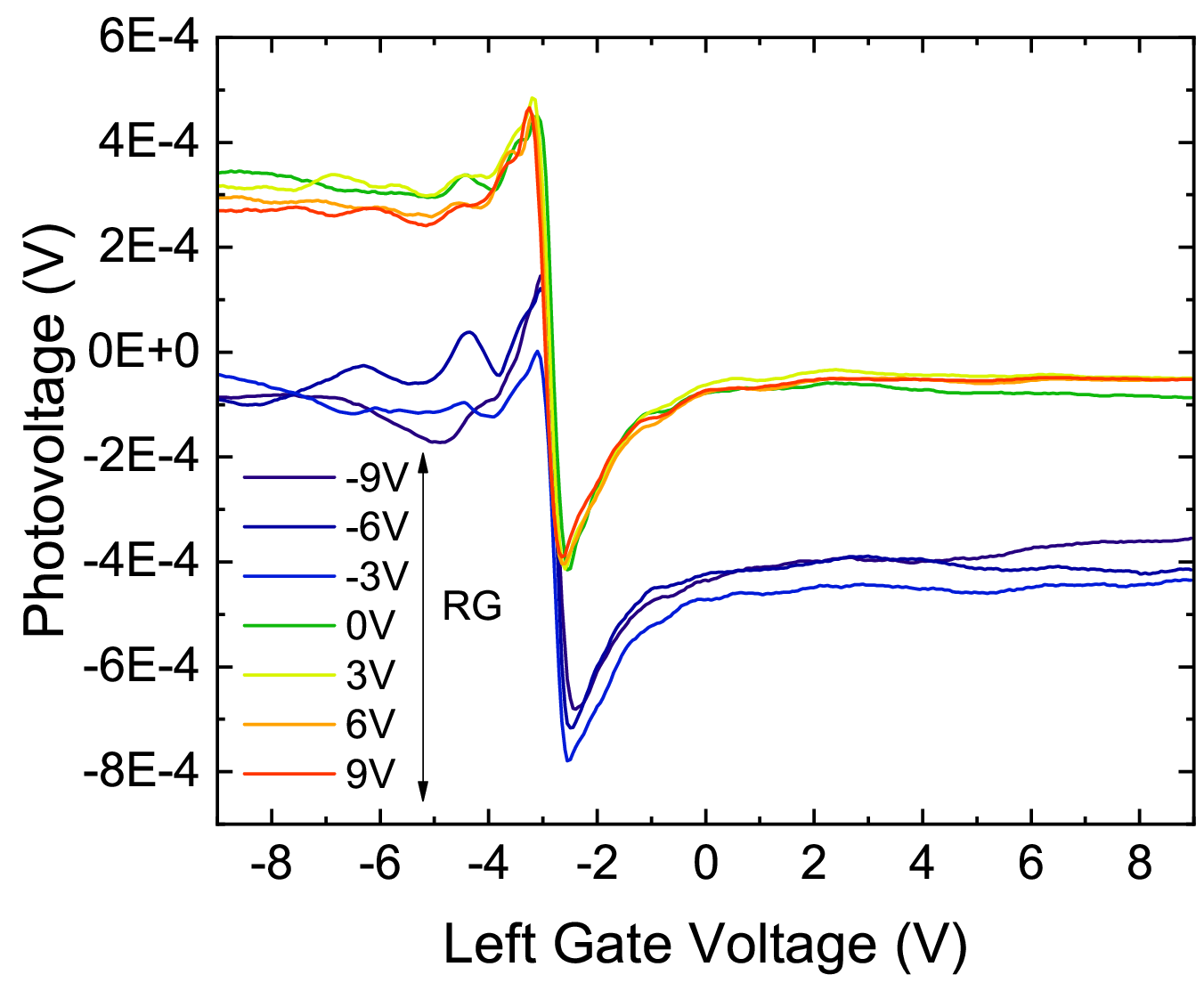}}
\caption{
Photovoltage $V_{\mathrm{ph}}$ as a function of left-gate voltage $V_{\mathrm{LG}}$ for several fixed right-gate voltages $V_{\mathrm{RG}}$ at $T = 6$~K.  
The curves correspond to $V_{\mathrm{RG}} = -9$~V to $+9$~V in 3~V steps, as indicated by the color scale.}
\label{fig:FigureS7}
\end{figure}

Supplementary Fig.~\ref{fig:FigureS7} shows the photovoltage $V_{\mathrm{ph}}$ measured as a function of left-gate voltage $V_{\mathrm{LG}}$ for several fixed values of the right-gate voltage $V_{\mathrm{RG}}$, recorded at $T = 6$~K under 2.5~THz illumination.  
The measurements reveal that the overall photoresponse remains qualitatively similar to the right-gate sweeps reported in the main text, exhibiting a pronounced polarity inversion around the charge neutrality point (CNP) and multiple smaller oscillatory features at gate voltages below the CNP. Compared to the right-gate dependence shown in Fig.~1 (h) of the main text, the magnitude of the photovoltage modulation with $V_{\mathrm{LG}}$ is reduced, and the detailed resonance structure becomes less pronounced. This is expected since the left gate primarily affects the overall Seebeck asymmetry between the two graphene regions, rather than directly tuning the carrier density and plasmonic resonance conditions in the active right-gated section.  

These results further corroborate the interpretation discussed in the main text: the right gate controls the formation and spectral position of the AGP resonances, while the left gate plays a secondary role by setting the background doping and thermoelectric contrast required for efficient PTE detection. This dataset is included here for completeness.

\section{Photovoltage Response at 1.8~THz}

\begin{figure*}[htbp!]
\centerline{\includegraphics[width=180mm]{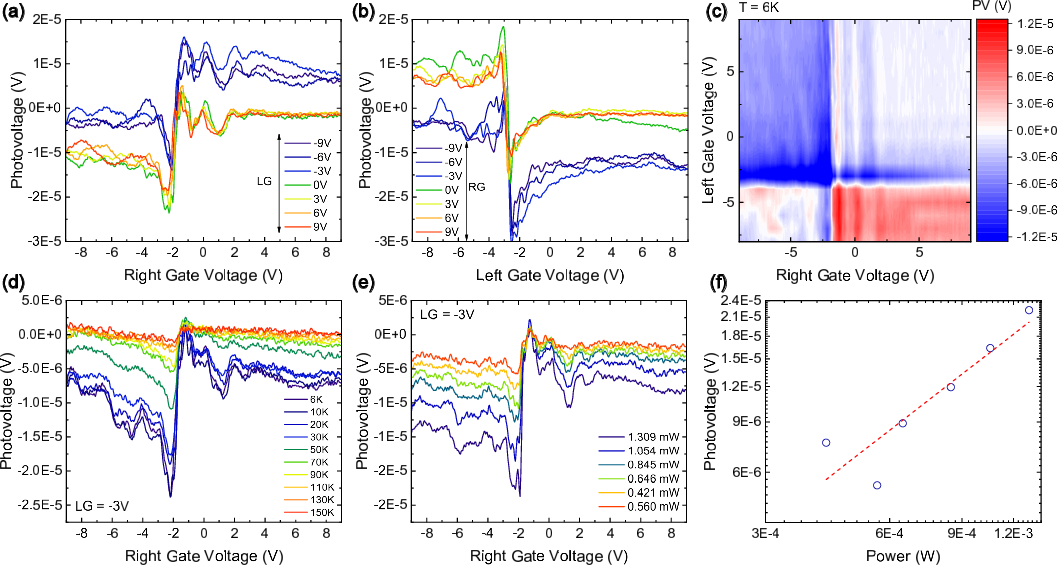}}
\caption{
Gate-, temperature-, and power-dependent photovoltage measurements at $f = 1.8$~THz.  
(a) Photovoltage $V_{\mathrm{ph}}$ as a function of right-gate voltage $V_{\mathrm{RG}}$ for several fixed left-gate voltages $V_{\mathrm{LG}}$.  
(b) Photovoltage as a function of $V_{\mathrm{LG}}$ for fixed $V_{\mathrm{RG}}$.  
(c) Two-dimensional map of $V_{\mathrm{ph}}(V_{\mathrm{LG}}, V_{\mathrm{RG}})$ at $T=6$~K.  
(d) and (e) Temperature- and power-dependent right-gate sweeps, respectively.  
(f) Peak photovoltage as a function of incident power, confirming a linear dependence.}
\label{fig:FigureS8}
\end{figure*}

Supplementary Fig.~\ref{fig:FigureS8} summarizes the device characterization under illumination at a lower excitation frequency of $f = 1.8$~THz.  
This measurement serves as a comparative study to the 2.5~THz data presented in the main text, illustrating the response of the same device when the antenna is operated off its designed resonance frequency.  
Overall, the observed signals confirm the persistence of the PTE response and the presence of gate-dependent modulations associated with plasmonic effects, although with reduced amplitude and increased noise. Panel~(a) displays the photovoltage $V_{\mathrm{ph}}$ as a function of right-gate voltage $V_{\mathrm{RG}}$ for several fixed left-gate voltages $V_{\mathrm{LG}}$.  
The polarity inversion around the charge neutrality point (CNP) remains visible, consistent with PTE-dominated behavior.  
Ripple-like oscillations and weaker resonant features appear on the electron-doped side, with broad maxima around $V_{\mathrm{RG}}\approx 0.2$~V and $1.2$~V.  
These features resemble the 1$^{\mathrm{st}}$ and 2$^{\mathrm{nd}}$ AGP resonances discussed at 2.5~THz, but the lower excitation frequency and off-resonant antenna coupling make their precise identification more uncertain. Panel~(b) shows complementary left-gate sweeps at different fixed $V_{\mathrm{RG}}$ values.  
The qualitative dependence is similar to that observed at higher frequency, although the modulation depth with $V_{\mathrm{LG}}$ is smaller, indicating weaker sensitivity of the PTE response to left-gate tuning under off-resonant excitation.  
Panel~(c) provides a two-dimensional color map of $V_{\mathrm{ph}}$ as a function of both gate voltages at $T=6$~K, reproducing the sixfold symmetry typical of dual-gated graphene detectors.  
The contrast is reduced compared to the 2.5~THz case, and the fine oscillatory structure is less pronounced due to the weaker coupling efficiency of the antenna at this frequency.

The temperature dependence is examined in panel~(d), where the amplitude of the photovoltage progressively decreases with increasing $T$, consistent with enhanced plasmon damping and reduced carrier mobility.  
The resonant-like modulations remain visible at low temperature but gradually fade above $\sim100$~K.  
Panel~(e) shows the effect of varying the incident THz power, confirming that the overall signal scales monotonically with excitation intensity, even in the off-resonant regime.  
Finally, panel~(f) plots the extracted photovoltage peak amplitude versus incident power on a linear scale, yielding an approximately linear relationship.  
This behavior confirms that the detector continues to operate in the linear response regime at 1.8~THz, despite the lower signal-to-noise ratio.

The results in Supplementary Fig.~\ref{fig:FigureS8} demonstrate that the device maintains a measurable PTE response and weak plasmonic modulations at 1.8~THz.  
However, the reduced photovoltage amplitude and increased noise level are consistent with operation away from the antenna resonance, as predicted by electromagnetic simulations (see main text Fig.~5(b)).  
These measurements further validate the robustness of the detection mechanism over a range of excitation frequencies, even though the plasmonic enhancement is optimized around 2.5~THz.

\end{document}